\documentclass[conference]{IEEEtran}

\usepackage{enumitem}
\usepackage{algorithm}
\usepackage{algorithmicx}
\usepackage[noend]{algpseudocode}
\usepackage{amsmath}
\usepackage{graphicx}
\usepackage{svg}
\usepackage{threeparttable}
\usepackage{booktabs}
\usepackage{pifont}
\usepackage{amsfonts}
\usepackage{textcomp}
\usepackage{xcolor}
\usepackage{multirow}
\usepackage{tabularx}
\usepackage[table]{xcolor}
\usepackage[most]{tcolorbox}
\usepackage{fvextra}
\usepackage{hyperref}
\usepackage{fancyvrb}
\usepackage[dvipsnames]{xcolor}

\ifCLASSINFOpdf
\else
\fi
\begin{document}

\title{VULPO: Context-Aware Vulnerability Detection via On-Policy LLM Optimization}

\author{
\IEEEauthorblockN{
Youpeng Li\IEEEauthorrefmark{1},
Fuxun Yu\IEEEauthorrefmark{2},
Weiliang Qi\IEEEauthorrefmark{1}
Xinda Wang\IEEEauthorrefmark{1}
}

\IEEEauthorblockA{\IEEEauthorrefmark{1}University of Texas at Dallas    \IEEEauthorrefmark{2}Microsoft}

\IEEEauthorblockA{
\texttt{\{youpeng.li, weiliang.qi, xinda.wang\}@utdallas.edu},\;
\texttt{fuxunyu@microsoft.com}
}
}

\maketitle

\begin{abstract}
Large language models (LLMs) have recently shown strong potential in vulnerability detection (VD). However, accurately detecting vulnerabilities in real-world repositories requires reasoning over complex contextual interactions. Existing LLM-based VD approaches remain limited because current datasets lack complete contextual information and high-quality reasoning supervision, while existing optimization methods primarily rely on coarse outcome-centric supervision signals that fail to model the vulnerability reasoning process.

To address these limitations, we first construct ContextVul, a new dataset that augments high-quality function-level vulnerability benchmarks with repository-level contextual information and curated vulnerability reasoning traces. Building upon ContextVul, we introduce a two-stage optimization framework consisting of lightweight cold-start supervised fine-tuning followed by vulnerability-adaptive on-policy optimization (VULPO). VULPO incorporates multidimensional rewards that jointly evaluate vulnerability identification, vulnerability-relevant localization, and causal reasoning quality, along with difficulty-adaptive reward scaling to mitigate reward hacking and improve RL effectiveness. Extensive experiments demonstrate the superiority of VULPO for context-aware VD. Our VULPO-4B, the first specialized vulnerability reasoning LLM, substantially outperforms existing VD baselines, improving Pairwise Pass@1 by 203\% relative to Qwen3-4B and achieving competitive performance against a 150× larger-scale LLM, DeepSeek-V3.1.

\end{abstract}

\IEEEpeerreviewmaketitle

\section{Introduction}

In real-world software repositories, determining whether a piece of code contains vulnerabilities requires considering its surrounding context. Security analysts typically rely on rich contextual information, such as surrounding logic, inter-procedural dependencies, and global or shared state. More importantly, this analysis goes beyond recognizing the presence of contextual elements. It requires reasoning about how these elements interact to determine whether a vulnerability can actually manifest or has already been mitigated.
However, existing learning-based vulnerability detection (VD) systems still struggle to replicate this human-like contextual reasoning capability, primarily due to fundamental limitations in both data and methodological design.

From a data perspective, existing VD datasets~\cite{52,51,26,21,1,54,5,53,55,2,50,28,56,30} rarely provide both complete vulnerability-relevant context and high-quality vulnerability reasoning chains, i.e., chain-of-thought (CoT) rationales. Early VD datasets are typically constructed by extracting vulnerable–fixed function pairs from patch commits and assigning classification labels. Such datasets largely restrict each sample to local function-level code, omitting the surrounding program context on which vulnerability manifestation often depends. In settings where such contextual information is missing, prior work has shown that most vulnerabilities cannot be accurately identified~\cite{12}. While later context-level datasets~\cite{2,50,28,56,30} attempt to address this limitation by incorporating broader code context, they still commonly suffer from incomplete or unreliable context extraction and, more importantly, provide only coarse-grained classification labels rather than high-quality reasoning chains that capture how vulnerabilities are analyzed. As a result, existing datasets not only limit a model’s ability to assess vulnerabilities based on code context, but also fail to provide effective supervisory signals for training context-aware VD models, thereby constraining their ability to learn the underlying reasoning mechanisms.

Methodologically, existing approaches are limited by training objectives that fail to effectively guide context-based vulnerability analysis. Many VD methods rely on simple binary or multi-class labels and use supervised fine-tuning (SFT) to predict whether code is vulnerable~\cite{7,16,20,39}. This outcome-oriented paradigm mainly learns a direct mapping from input code to labels, without explicitly modeling the contextual conditions under which vulnerabilities arise. Thus, models tend to rely on superficial statistical features or local patterns instead of performing thorough contextual reasoning~\cite{12}.
Recent works adopt preference optimization methods, such as Direct Preference Optimization (DPO~\cite{22}) and Odds-Ratio Preference Optimization (ORPO~\cite{23}), by constructing label-derived chosen/rejected response pairs to improve discrimination between vulnerable and patched code~\cite{36,37}. However, these approaches exhibit a fundamental limitation: they encourage post-hoc rationalization, where models produce superficially plausible explanations that align with the given label, rather than to derive conclusions through genuine context-based reasoning. In essence, such preference modeling remains at the level of outcome-based comparison and does not explicitly model or constrain the vulnerability analysis process. Moreover, its reliance on statically constructed preference data limits the diversity of contextual interactions and reasoning paths that models can learn. Consequently, these methods still fall short of guiding models toward context-aware reasoning.
Reinforcement learning (RL)-based approaches attempt to optimize model behavior through dynamic interaction, but they typically rely on classification correctness as the primary reward signal~\cite{29}. This outcome-centric reward design lacks fine-grained constraints on reasoning process, allowing models to achieve high rewards by exploiting shortcuts (i.e., reward hacking), without genuinely modeling vulnerability-relevant contextual semantics or the underlying reasoning mechanisms.

To address the above limitations, we first construct ContextVul, a new dataset specifically designed to support context-aware VD. Unlike prior datasets that provide limited contextual coverage and lack reasoning supervision, ContextVul enriches high-quality function-level vulnerability benchmarks~\cite{1,30} with repository-level contextual information. It uses clang-based builds~\cite{57}, with Joern~\cite{31} as a fallback mechanism, to enable scalable extraction of inter-procedural and cross-file dependencies. Moreover, ContextVul provides high-quality vulnerability reasoning traces generated by a teacher LLM and filtered through LLM-as-a-judge-assisted rejection sampling, thereby offering high-quality and structured supervision beyond outcome-level labels.

Building on ContextVul, we propose a two-stage policy optimization framework tailored for context-aware VD. In the first stage, we perform lightweight cold-start initialization via SFT on the curated vulnerability reasoning dataset, enabling the model to generate informative rollouts for subsequent on-policy RL training. In the second stage, we introduce \underline{Vul}nerability-adaptive \underline{P}olicy \underline{O}ptimization (VULPO), a tailored RL method that shifts training from outcome-centric to process-aware supervision. Specifically, VULPO incorporates multidimensional reward signals evaluated by an LLM-based judge through rubric-guided assessments grounded in vulnerability metadata. These rewards jointly evaluate vulnerability identification, vulnerability-relevant localization, and causal reasoning quality. To further improve training stability and learning efficiency, we design a difficulty-adaptive reward scaling mechanism at both label and sample levels. Label-level reward scaling addresses the inherent difficulty asymmetry between positive and negative predictions, which may otherwise cause the model to exploit easier negative predictions and lead to reward hacking. Sample-level reward scaling, in turn, amplifies rewards for correct rollouts on more challenging samples, encouraging the model to explore more diverse solution paths and improving learning efficiency.

Extensive evaluation evidences the effectiveness of the proposed framework from multiple perspectives. On the ContextVul benchmark, VULPO-4B achieves a P-Pass@1 of 22.60\%, P-Pass@8 of 53.91\%, and F1 score of 40.09\%, substantially outperforming both existing optimization-based VD methods and several frontier reasoning LLMs with significantly larger parameter scales. In particular, despite using only a 4B backbone model, VULPO surpasses DeepSeek-V3.1, DeepSeek-R1-0528, and GPT-OSS-120B on multiple evaluation metrics, demonstrating the effectiveness and scalability of process-centric RL optimization for context-aware VD.

In addition to overall performance improvements, extensive analyses further validate the design motivations behind VULPO: (i) Lightweight cold-start SFT is critical for balancing rollout correctness and exploration during RL training. (ii) Label-difficulty-aware reward scaling effectively mitigates reward hacking caused by prediction difficulty asymmetry between vulnerable and non-vulnerable samples. (iii) Sample-difficulty-aware reward scaling improves learning efficiency by encouraging exploration of challenging vulnerability reasoning trajectories. (iv) Process-aware multidimensional rewards significantly improve quality and faithfulness of vulnerability reasoning compared with coarse outcome-based supervision. (v) Hierarchical correctness analysis demonstrates that conventional binary-label or CWE-level evaluations substantially overestimate true VD capability, highlighting necessity of root-cause-level supervision and evaluation.

We summarize our contributions as follows:
\begin{itemize}[leftmargin=10pt]
    \item We construct ContextVul, a new context-aware VD dataset for C/C++ that augments high-quality function-level vulnerability benchmarks with repository-level contextual information and curated vulnerability reasoning traces.
    \item We propose VULPO, a process-centric RL framework for context-aware VD that optimizes reasoning behaviors via multidimensional rewards evaluating vulnerability identification, localization, and causal reasoning quality.
    \item We design a difficulty-adaptive reward scaling mechanism at both the label and sample levels to mitigate reward hacking caused by prediction difficulty asymmetry and improve learning efficiency during RL training.
    \item Extensive experiments highlight that VULPO-4B,  first specialized vulnerability reasoning LLM, achieves superior performance on ContextVul while exhibiting strong generalization capability on out-of-distribution vulnerabilities.
    \item The artifact of VULPO is available \href{https://huggingface.co/collections/Leopo1d/vulpo}{here}.
\end{itemize}

\section{Motivation and Overview}\label{sec:mtv}
\begin{figure}[h]
\centering
\includegraphics[width=0.9\columnwidth]{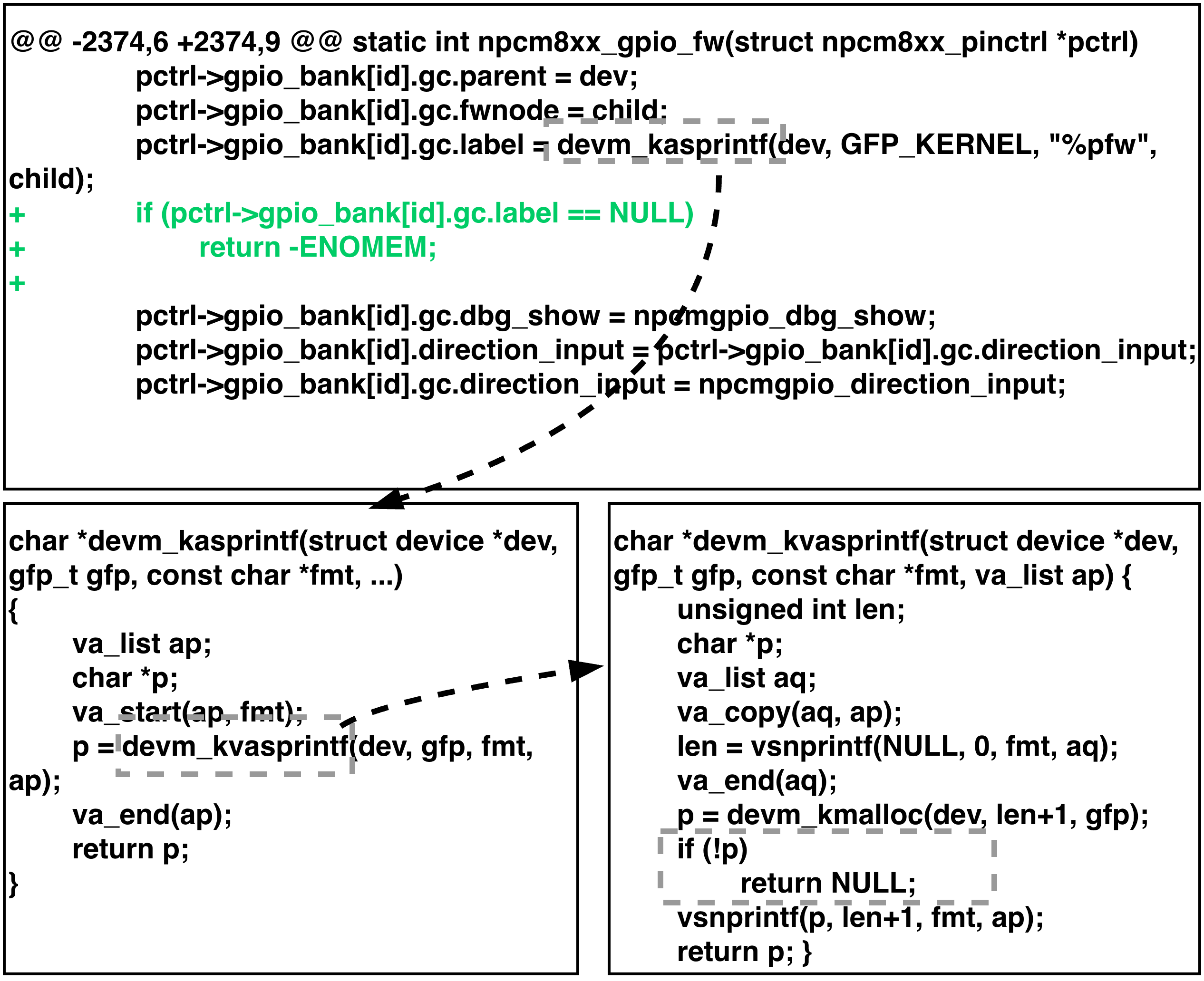}
\caption{Cross-Function Context Reveals a Missing Null Check} 
\label{fig:motivation}
\end{figure}

\subsection{Necessity of Context in Vulnerability Detection} Figure~\ref{fig:motivation} shows a real-world motivating example from CVE-2025-21982, a null pointer dereference vulnerability. In the patch shown at the top, the calling function \texttt{\small npcm8xx\_gpio\_fw()} invokes \texttt{\small devm\_kasprintf()} and uses its return value without a null check. Although \texttt{\small devm\_kasprintf()} itself does not explicitly indicate a possible null return in its signature, it further calls \texttt{\small devm\_kvasprintf()}, a well-established kernel helper whose implementation may return \texttt{\small NULL} when memory allocation fails. Therefore, the vulnerability is not locally observable from the caller alone. Instead, it is an inter-procedural property that can only be revealed through call-graph context.
When prompting with only the caller body, LLMs lack the information needed to infer that the callee may return null, and models such as Qwen3-4B consistently predict such sites as safe. Supplying the implementation of \texttt{\small devm\_kasprintf()} alongside its callee resolves this ambiguity, enabling the model to correctly identify the missing null check as a defect. Therefore, cross-function context is essential for detecting vulnerabilities that are invisible under function-level analysis.

\subsection{Limitations of Outcome-Centric RL for Context-Aware VD}

\begin{figure}[h]
\centering
\includegraphics[width=0.85\columnwidth]{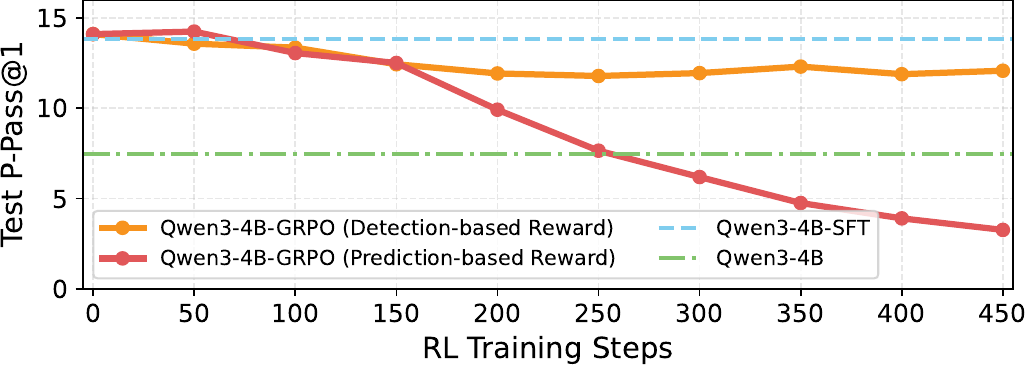}
\caption{Coarse Outcome-Based Rewards Limit RL Training Effectiveness} 
\label{fig:motivation2}
\end{figure}

In VD, a model's ability to reason over contextual information directly determines the correctness of its final prediction. However, existing RL-based VD methods~\cite{29} predominantly adopt outcome-centric reward designs that focus only on the correctness of final classification results, while neglecting the validity of the underlying reasoning process. This “heavy on outcome, light on process” paradigm can reward cases where flawed reasoning still yields correctly guessed answers, introducing spurious reward signals and disrupting effective RL training.
To empirically validate this, we conduct a preliminary study examining the impact of reward design on RL training. Specifically, we fine-tune Qwen3-4B on ContextVul to obtain an initial SFT model, which is then used to initialize the policy model in the RL stage. We consider two representative outcome-based reward designs: (i) a detection-based reward, which evaluates the correctness of binary vulnerability classification, (ii) a prediction-based reward, which evaluates the correctness of predicted CWE types. Both rewards are used to guide Group Relative Policy Optimization (GRPO) training.

As shown in Figure~\ref{fig:motivation2}, outcome-based rewards significantly restrict effective exploration during RL training, resulting in final performance worse than the SFT baseline despite additional computational overhead. This degradation can be attributed to the lack of process-level supervision. Under detection-based rewards, rollouts that produce correct final predictions but rely on incorrect reasoning are still rewarded, causing misleading training signals. Under prediction-based rewards, the problem becomes even more severe. Since predicting the correct CWE type is much more difficult than predicting the absence of a vulnerability, the imbalance in prediction difficulty across labels encourages reward hacking, where the model learns to predict all samples as safe to maximize rewards, resulting in severe model collapse.

These findings reveal that coarse-grained outcome-based rewards fail to impose fine-grained constraints on the vulnerability reasoning process, highlighting the need for a context-aware, process-centric RL framework for VD.

\subsection{System Design Overview}

\begin{figure*}[h]
\centering
\includegraphics[width=0.8\textwidth]{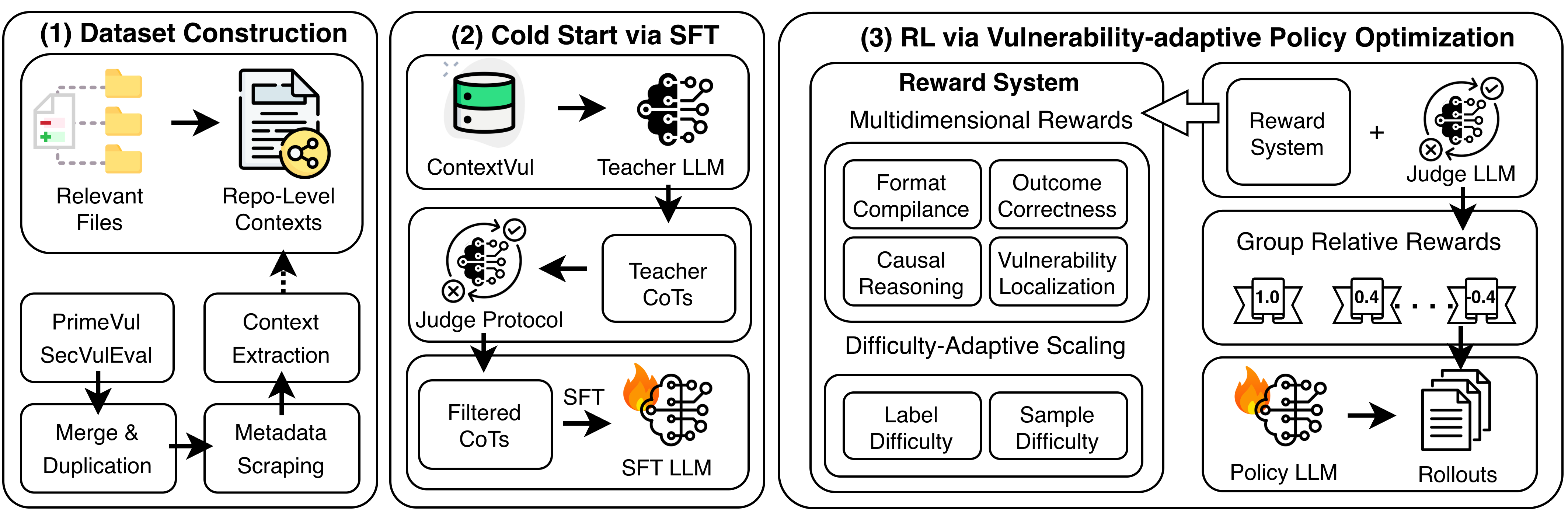}
\caption{System Overview} 
\label{fig:VULPO}
\end{figure*}

Figure~\ref{fig:VULPO} presents an overview of our system architecture, which consists of four tightly integrated components: construction of a context-aware VD dataset, a cold-start stage for building fundamental VD capabilities, a VULPO-based RL stage, and an LLM-based judge protocol for both reward assessment and performance evaluation. Together, these components form a principled pipeline for developing specialized reasoning LLMs for context-aware VD.

\section{Context-Aware VD Dataset Construction}\label{sec:con}

\begin{table}[h]
\caption{Comparison of Representative VD Datasets}
\label{tab:vd_datasets}
\centering
\resizebox{\columnwidth}{!}{
\renewcommand{\arraystretch}{1.0}
\begin{tabular}{ccc}
\toprule
\textbf{Vulnerability Detection Dataset} & \textbf{Context Integrity} & \textbf{Reasoning Supervision} \\ \midrule
\begin{tabular}[c]{@{}c@{}}BigVul\cite{52}, CVEFixes\cite{51}, SVEN\cite{26}, DiverseVul\cite{21} \\ PrimeVul\cite{1}, MegaVul\cite{54}, SecLLMHolmes\cite{5} \\ CleanVul\cite{53}, BenchVul\cite{55}, TitanVul\cite{55}  \end{tabular} & \ding{109} & \ding{55} \\ \midrule
\begin{tabular}[c]{@{}c@{}}ReposVul\cite{2}, JitVul\cite{50}, CORRECT\cite{28}\\ SECODEPLT\cite{56}, SecVulEval\cite{30}\end{tabular} & \ding{119} & \ding{55} \\ \midrule
ContextVul (Ours) & \ding{108} & \ding{51} \\ \bottomrule
\end{tabular}
}
\end{table}

Most existing VD datasets provide only function-level code snippets, omitting the repository-level information necessary for context-aware VD~\cite{52,51,26,21,1,54,5,53,55}. As a result, models are forced to reason over incomplete program semantics and cannot capture the cross-file interactions, inter-procedural dependencies, and global state behaviors that often determine whether a vulnerability can manifest or has been mitigated. Therefore, LLMs trained under such restricted settings may produce false positives by overlooking external sanitization or validation routines, or false negatives by failing to capture missing trigger conditions. Moreover, evaluations based on isolated functions poorly reflect real-world detection performance in large codebases, where contextual signals critically influence vulnerability manifestation~\cite{12}.

Although some prior works incorporate contextual information into VD, they have several limitations as summarized in Table~\ref{tab:vd_datasets}: (i) Datasets such as JitVul~\cite{50}, ReposVul~\cite{2}, and SecVulEval~\cite{30} are claimed as context-level datasets, yet many samples contain little or no meaningful context. (ii) CORRECT~\cite{28} relies on ground-truth vulnerability locations (e.g., patch diffs) to extract context, which is unrealistic because such information is unavailable in real-world detection scenarios. (iii) Several context-level datasets (e.g., SECODEPLT~\cite{56} and CORRECT~\cite{28}) use shared contexts for both vulnerable and patched code, failing to capture contextual elements newly introduced by the fix. (iv) Existing datasets all provide only classification labels without reasoning supervision signals, making them unsuitable for training specialized reasoning LLMs for VD that require high-quality CoTs.

To bridge this gap, we build {ContextVul}, a context-aware VD dataset for C/C++ that enriches target functions with their surrounding contextual information. ContextVul is built upon two carefully curated and labeled VD datasets, PrimeVul~\cite{1} and SecVulEval~\cite{30}. To mitigate the risk of potential unknown vulnerabilities in unchanged functions within patch commits, we retain only paired samples, where each pre-patch vulnerable function is matched with its corresponding post-patch fixed version and associated with a labeled vulnerability. The data construction of ContextVul includes the following two stages.

\noindent\textbf{Data Preprocessing.}
Due to missing metadata or expired links for some commits in the two source datasets, we supplement the corresponding vulnerability information through data scraping and manual verification to include commit links, commit messages, commit dates, changed files and functions involved in each commit, repository names, project names, CWE IDs, CVE IDs, and CVE descriptions. We then merge the two datasets and remove duplicate samples to prevent data leakage, ensuring that the same sample does not appear in both the training and test sets. We also remove comments from the code to avoid leakage from comments containing vulnerability-related information such as CVE IDs, enabling a more accurate evaluation of model performance. 
    
\noindent\textbf{{Context Extraction.}}
To capture the semantic context surrounding each target function, we implement an automated extraction pipeline built on top of Claude Code~\cite{58} and based on libclang, the C interface to the Clang compiler frontend~\cite{57}. Given a tuple consisting of a repository, commit, file path, and target function, the pipeline checks out the corresponding source version, constructs a compilation database when feasible, and performs AST-level analysis to extract contextual elements, including functions within two hops in the call graph, macros, type definitions, global variables, and include directives. When clang-based analysis fails due to incomplete build configurations or unsupported compilation environments, we employ Joern~\cite{31} as a fallback static analysis backend. Joern constructs a Code Property Graph (CPG) that integrates abstract syntax, control flow, and data-flow representations, enabling robust extraction of structural and semantic context.
This hybrid extraction strategy improves the robustness and coverage of contextual recovery across heterogeneous repositories. 
To assess the quality of the context extraction pipeline, two authors independently inspect 100 randomly sampled instances and confirm that the extracted contextual elements are correct in over 95\% of the cases.

\section{Cold Start via Supervised Fine-Tuning}\label{sec:sft}

Due to the complexity of VD, directly applying RL to relatively small-scale LLMs often leads to poor training effectiveness, as the model fails to generate sufficiently informative rollouts~\cite{25}. To solve this, we perform a lightweight cold-start initialization using SFT to improve the model’s initial capability before RL training.

\noindent\textbf{Reasoning Data Curation.}
To construct a high-quality reasoning dataset for SFT, we employ DeepSeek-R1-0528~\cite{25}, one of the strongest open-source LLMs available during our study, as the teacher LLM to generate vulnerability reasoning CoTs from context-level inputs. For each input, we sample multiple rollouts, eight in our implementation, and apply an LLM-based judge protocol to filter out incorrect or low-quality reasoning traces through rejection sampling.

\noindent\underline{LLM-based Judge Protocol.}
The judge evaluates the quality of the teacher model's vulnerability analysis based on vulnerability-related metadata associated with each sample. For instance, using the CVE description and patch commit message as references, it assesses whether the model correctly identifies the root cause of the target vulnerability or correctly infers its absence by recognizing existing safeguards. We retain only reasoning traces that satisfy the evaluation criteria. Detailed prompts are provided in Appendix~\ref{sec:specification}.

\noindent\textbf{Supervised Fine-Tuning.}
We conduct SFT on the curated reasoning dataset to initialize the policy LLM. We intentionally use only a small number of training epochs, as our analysis in Section~\ref{sec:cold} shows that excessive SFT weakens the model's exploration ability during RL training and degrades final performance.

\section{Reinforcement Learning with Vulnerability-Adaptive Policy Optimization}\label{sec:vulpo}

\subsection{{Problem Formulation}}\label{sec:grpo}

VULPO is built upon GRPO~\cite{24} where the optimization problem is formulated by maximizing the following objective:
\begin{equation}\label{eq:grpo}
\begin{split}
\mathcal{J}_{GRPO}(\theta) &= \mathbb{E}_{q \sim P(q), \{o_i\}_{i=1}^G \sim \pi_{\theta_{old}}(O|q)} \Bigg[ \frac{1}{G} \sum_{i=1}^G \frac{1}{|o_i|} \sum_{t=1}^{|o_i|} \\
&\quad \bigg( \min \Big[ \tfrac{\pi_{\theta}(o_{i,t}|q_i, o_{i,<t})}{\pi_{\theta_{old}}(o_{i,t}|q_i, o_{i,<t})} \hat{A}_{i,t}, \\
&\qquad \text{clip}\big(\tfrac{\pi_{\theta}(o_{i,t}|q_i, o_{i,<t})}{\pi_{\theta_{old}}(o_{i,t}|q_i, o_{i,<t})}, 1-\epsilon, 1+\epsilon\big) \hat{A}_{i,t} \Big] \\
&\quad - \beta D_{KL} \left[ \pi_{\theta} || \pi_{ref} \right] \bigg) \Bigg].
\end{split}
\end{equation}

\noindent Assume that, in the VD task, the input query $q$ is sampled from the VD dataset with data distribution $P(q)$. Subsequently, $G$ responses $\{o_i\}_{i=1}^G$ are sampled by feeding $q$ into the old policy model $\pi_{\theta_{old}}$. The reward model then scores all group responses $\{o_i\}_{i=1}^G$ to obtain the group rewards $\{r_i\}_{i=1}^G$, denoted as $\mathbf{r}$, from which the advantage is computed via group-wise normalization and shared across all time steps (i.e., $\hat{A}_{i,t} = \hat{A}_i = \frac{r_i - \text{mean}(\mathbf{r})}{\text{std}(\mathbf{r})}$).
In the pure on-policy setting of GRPO, the policy model $\pi_\theta$ is updated only once per batch, such that $\pi_\theta = \pi_{\theta_{old}}$. Under this setting, the objective in Equation~\ref{eq:grpo} simplifies to:
\begin{equation}\label{eq:grpo2}
\begin{split}
\mathcal{J'}_{GRPO}(\theta) &= \mathbb{E}_{q \sim P(q), \{o_i\}_{i=1}^G \sim \pi_{\theta_{old}}(O|q)} \bigg[ \frac{1}{G} \sum_{i=1}^G \frac{1}{|o_i|} \sum_{t=1}^{|o_i|} \\
&\quad \Big(\hat{A}_{i,t} - \beta D_{KL} \big[ \pi_{\theta} || \pi_{ref} \big] \Big) \bigg].
\end{split}
\end{equation}


\noindent\textbf{Limitations of Outcome-Based Rewards.}
In context-aware VD, each response $o_i$ typically contains both a reasoning trace, which performs context-based vulnerability analysis, and a final prediction indicating whether the target vulnerability exists. However, in existing RL-based approaches for VD~\cite{29}, the reward $r_i$ assigned to each response $o_i$ is typically outcome-based, reflecting only the correctness of the final prediction. Thus, the corresponding advantage $\hat{A}_i$ provides only a coarse-grained signal and fails to distinguish among responses with correct predictions but varying reasoning quality.

\begin{table*}[h]
\caption{Process-Aware Multidimensional Reward Design}
\label{tab:reward}
\centering
\renewcommand{\arraystretch}{0.85}
\begin{threeparttable}
\scriptsize
\resizebox{0.85\textwidth}{!}{
\begin{tabular}{clll}
\toprule
\textbf{Dimension} & \multicolumn{1}{c}{\textbf{Objective}} & \multicolumn{1}{c}{\textbf{Evaluation Basis}} & \multicolumn{1}{c}{\textbf{Scoring Rubric}} \\ \midrule
\begin{tabular}[c]{@{}c@{}}Outcome\\ Correctness\end{tabular} & \begin{tabular}[c]{@{}l@{}}Determine whether the response\\ warrants process-level evaluation.\end{tabular} & \begin{tabular}[c]{@{}l@{}}Vulnerability metadata:\\ CVE description and patch commit.\end{tabular} & \begin{tabular}[c]{@{}l@{}}+1.0: Correct prediction.\\ -1.0: Incorrect prediction.\end{tabular} \\ \midrule
\begin{tabular}[c]{@{}c@{}}Vulnerability\\ Localization\end{tabular} & \begin{tabular}[c]{@{}l@{}}Assess whether the model identifies\\ vulnerability-relevant code regions.\end{tabular} & Patch diff: removed/added code block & \begin{tabular}[c]{@{}l@{}}1.0: Correct localization.\\ 0.5: Partially correct.\\ 0.0: Incorrect or missing.\end{tabular} \\ \midrule
\begin{tabular}[c]{@{}c@{}}Causal\\ Reasoning\end{tabular} & \begin{tabular}[c]{@{}l@{}}Evaluate the quality of\\ root-cause and impact analysis.\end{tabular} & \begin{tabular}[c]{@{}l@{}}Semantic alignment with CVE description\\ or patch commit message.\end{tabular} & \begin{tabular}[c]{@{}l@{}}1.0: Aligned.\\ 0.5: Partially aligned.\\ 0.0: Not aligned.\end{tabular} \\ \midrule
\begin{tabular}[c]{@{}c@{}}Format\\ Adherence\end{tabular} & \begin{tabular}[c]{@{}l@{}}Ensure outputs follow the required\\ reasoning-answer structure.\end{tabular} & \begin{tabular}[c]{@{}l@{}}String match:\\ \texttt{<think>...</think>[answer]}\end{tabular} & \begin{tabular}[c]{@{}l@{}}+0.0: Valid format.\\ -0.2: Invalid format.\end{tabular} \\ \bottomrule
\end{tabular}
}
\end{threeparttable}
\end{table*}

\subsection{{Vulnerability-Adaptive Reward System}}
To address the above issue, we propose a vulnerability-adaptive reward system consisting of two key components: (i) a process-aware multidimensional reward design that jointly evaluates the reasoning process and the final outcome, and (ii) a difficulty-adaptive reward scaling mechanism operating at both the label and sample levels to improve training stability and learning efficiency during RL.

\noindent\textbf{Process-Aware Multidimensional Reward Design.} As demonstrated in Section~\ref{sec:mtv}, outcome-centric rewards that rely solely on final prediction correctness can lead to performance degradation and even model collapse. In contrast, a helpful VD model should provide structured vulnerability analysis, including identifying vulnerability-relevant code regions (where), diagnosing the root cause (why), explaining its impact (how), and producing a faithful final prediction. Motivated by this, we design a process-aware reward mechanism that evaluates both the reasoning process and the final outcome, providing fine-grained supervision for vulnerability analysis.

\noindent{\underline{\ding{182} Vulnerability Identification (Outcome Correctness).}} We treat outcome correctness as a gating signal that determines whether deeper evaluation of the reasoning process is necessary. Specifically, if the model produces an incorrect prediction, either failing to identify a vulnerability in vulnerable code or incorrectly predicting the already-fixed vulnerability in patched version, a penalty of $-1$ is assigned, and no further reward evaluation is performed. Otherwise, a reward of $+1$ is assigned, and the response proceeds to subsequent evaluation stages.
Instead of relying on CWE labels as the ground truth, we leverage vulnerability metadata such as CVE descriptions and patch commit messages to determine the target vulnerability. This design is motivated by two observations. First, inconsistencies in CVE-CWE mappings across vulnerability databases, along with the hierarchical structure of the CWE taxonomy, make exact CWE matching unreliable and misleading~\cite{46}. Second, we observe that models often correctly identify vulnerabilities while producing inaccurate CWE labels, or correctly predict general vulnerability categories but fail to provide accurate root-cause analysis aligned with the target CVE.

\noindent{\underline{\ding{183} Vulnerability-Relevant Localization.}} This reward distinguishes responses that correctly localize vulnerability-relevant code from those that rely on superficial pattern matching. We use patch commit differences as references. For vulnerable code, correct localization is determined by whether the model identifies the vulnerability-relevant code region (i.e., removed code in the patch). For patched code, localization is evaluated based on whether the model correctly explains how added code prevents the vulnerability. The evaluation is categorized as \texttt{\small CORRECT}, \texttt{\small PARTIALLY CORRECT}, and \texttt{\small INCORRECT}, corresponding to reward scores of $1.0$, $0.5$, and $0.0$, respectively.

\noindent{\underline{\ding{184} Causal Reasoning Quality.}} This reward incentivizes the model to produce semantically grounded and well-reasoned vulnerability analysis. For vulnerable code, reasoning quality is evaluated based on its semantic alignment with vulnerability description, which typically captures root cause and impact. For patched code, reasoning quality is assessed by its alignment with patch commit message, which often explains how the fix mitigates vulnerability. The evaluation is categorized as \texttt{\small CORRECT}, \texttt{\small PARTIALLY CORRECT}, and \texttt{\small INCORRECT}, corresponding to reward scores of $1.0$, $0.5$, and $0.0$, respectively.

\noindent{\underline{\ding{185} Format
Adherence.}} We use a simple string match to check if the model's output follows the specific reasoning-answer template (i.e., \textless think\textgreater [reasoning]\textless /think\textgreater [answer]). Since many reasoning LLMs (e.g., Qwen3) already have strong format compliance, we apply a penalty of $-0.2$ only when the model’s output fails to meet the format.

In this way, our vulnerability-adaptive reward system provides process-aware evaluation by assessing both the reasoning process and the final outcome from multiple dimensions, rather than relying solely on outcome correctness. Section~\ref{sec:ablation_study} will confirm that the proposed reward design provides reliable supervision signals for RL training.

\subsection{LLM Judges as Reward Models}\label{sec:reward_judge}

To operationalize the proposed process-aware reward design, we require a reward model capable of evaluating both the reasoning process and final outcome of the model analysis. In VD, training a parametric reward model is particularly challenging due to the scarcity of high-quality annotated data for reasoning processes and the difficulty of obtaining reliable supervision signals beyond coarse-grained labels. As a result, learned reward models often suffer from limited generalization and may fail to provide accurate feedback for complex vulnerability analysis.
To address this limitation, we employ an LLM-based judge as the reward model. Specifically, we leverage publicly available vulnerability information from the National Vulnerability Database (NVD)~\cite{nvd}, including CVE descriptions, patch commit messages, and code differences, as ground-truth references.

\begin{figure*}[h]
\centering
\includegraphics[width=0.85\textwidth]{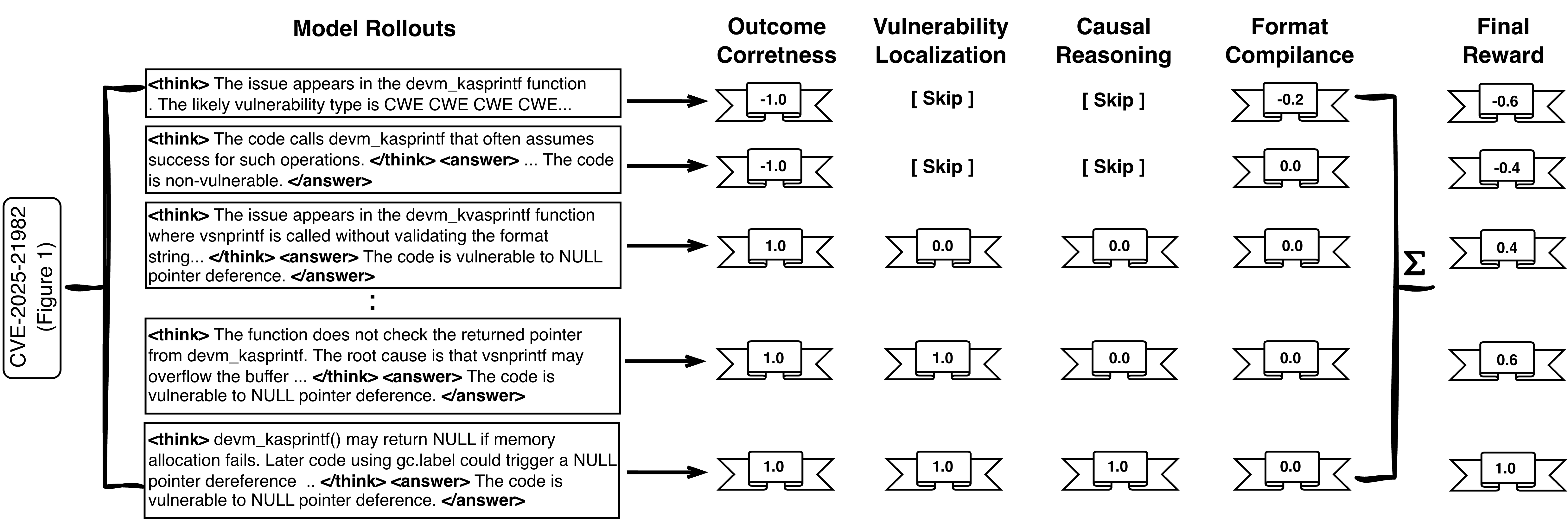}
\caption{Workflow of Process-Aware Multidimensional Reward Evaluation}
\label{fig:reward_pipeline}
\end{figure*}

Given a model response and the corresponding vulnerability metadata, the judge LLM evaluates the output across multiple dimensions based on predefined guidelines. Figure~\ref{fig:reward_pipeline} presents an example of the multidimensional reward evaluation process, including outcome correctness gating, process-level reasoning assessment, and final reward aggregation. For each dimension, the judge produces both a categorical judgment and a corresponding score. Table~\ref{tab:reward} summarizes the reward design and scoring rubrics, and the full prompts are provided in Appendix~\ref{sec:specification}.
In this way, the judge LLM translates structured vulnerability knowledge into fine-grained reward signals, enabling process-centric supervision over the vulnerability reasoning process during RL training.

\noindent\textbf{Reward Aggregation.}
To obtain the final reward, we aggregate the scores from different dimensions using a weighted formulation. Specifically, we define the total reward as:
\begin{equation}
R = w_{\text{out}} \cdot r_{\text{out}}
+ \mathbf{1}(r_{\text{out}} > 0) \cdot
\left(
w_{\text{loc}} \cdot r_{\text{loc}}
+ w_{\text{reason}} \cdot r_{\text{reason}}
\right)
+ r_{\text{format}}.
\end{equation}
Here, $r_{\text{out}}$, $r_{\text{loc}}$, and $r_{\text{reason}}$ denote the scores for outcome correctness, vulnerability-relevant localization, and causal reasoning quality, respectively, while $w_{\text{out}}$, $w_{\text{loc}}$, and $w_{\text{reason}}$ are their corresponding weights, and $\mathbf{1}(\cdot)$ is an indicator function that activates process-level rewards only when the outcome prediction is correct (i.e., $r_{\text{out}} > 0$), reflecting the gating role of outcome correctness. In addition, the format reward $r_{\text{format}}$ is treated as an auxiliary penalty term that is only applied when the model output violates the required reasoning-answer format, and does not participate in the weighted aggregation.
In our implementation, we set $w_{\text{out}} = 0.4$, $w_{\text{loc}} = 0.2$, and $w_{\text{reason}} = 0.4$, corresponding to a 2:1:2 weighting scheme. This design reflects the process-aware nature of vulnerability analysis. While correct prediction is necessary, localization and reasoning quality are both critical for ensuring faithful and actionable analysis. In practice, we observe that models may produce correct predictions without accurately localizing the vulnerability or identifying its root cause. By assigning substantial weight to causal reasoning quality, the reward design encourages the model to produce structured and context-aware analysis rather than relying on superficial patterns.

Overall, this process-aware multidimensional reward design enables the judge LLM to distinguish not only between vulnerability analyses with correct and incorrect outcomes, but also among varying qualities of reasoning, thereby providing more stable and informative supervision signals for RL training.

\subsection{{Difficulty-Adaptive Reward Scaling}}

In RL training, assigning uniform rewards to all samples can lead to inefficient learning and undesirable optimization behavior. Samples that the model already predicts correctly should intuitively contribute less to learning, whereas more challenging samples should be emphasized to encourage effective exploration.
This issue in VD is amplified by inherent difficulty variations across both labels and individual samples.  Particularly, predicting code as non-vulnerable is generally easier than correctly identifying vulnerabilities and analyzing their root causes. Here, we propose two complementary difficulty-aware reward scaling strategies to mitigate reward hacking and improve learning efficiency.

\noindent{\textbf{Label Difficulty-Aware Reward Scaling.}} As observed in Section~\ref{sec:mtv}, in LLM-based VD, if equal rewards are assigned to true positives and true negatives, the model tends to exploit easier negative predictions to maximize rewards, leading to an increase in false negatives despite a balanced data distribution.
To address this, we assign higher rewards and penalties to vulnerable samples. We introduce a label difficulty weight $w_l$ ($w_l > 1$) applied to vulnerable samples to adjust reward magnitude accordingly.
The choice of $w_l$ depends on the intrinsic difficulty of the VD task and the model’s initial prediction bias (e.g., the accuracy gap between positive and negative predictions). We analyze the impact of $w_l$ in Section~\ref{sec:labeldiff}.

\noindent{\textbf{Sample Difficulty-Aware Reward Scaling.}} LLMs often exhibit varying performance across vulnerability samples due to differences in code complexity, vulnerability types, and prior exposure during pre-training. However, explicitly estimating sample difficulty is challenging in practice.
Inspired by GRPO (Section~\ref{sec:grpo}), we approximate sample difficulty using the model’s on-policy behavior. For each sample, the policy generates $G$ rollouts per training step. A higher proportion of correct responses indicates that the sample is relatively easy, whereas a lower proportion suggests higher difficulty.
Based on this, we introduce a sample difficulty weight $w_s$ ($w_s > 0$), which is inversely proportional to the proportion of correct responses. This design amplifies rewards for harder samples, encouraging the model to focus on challenging cases and improving learning efficiency.

\begin{figure}[h]\centering
\includegraphics[width=0.38\textwidth]{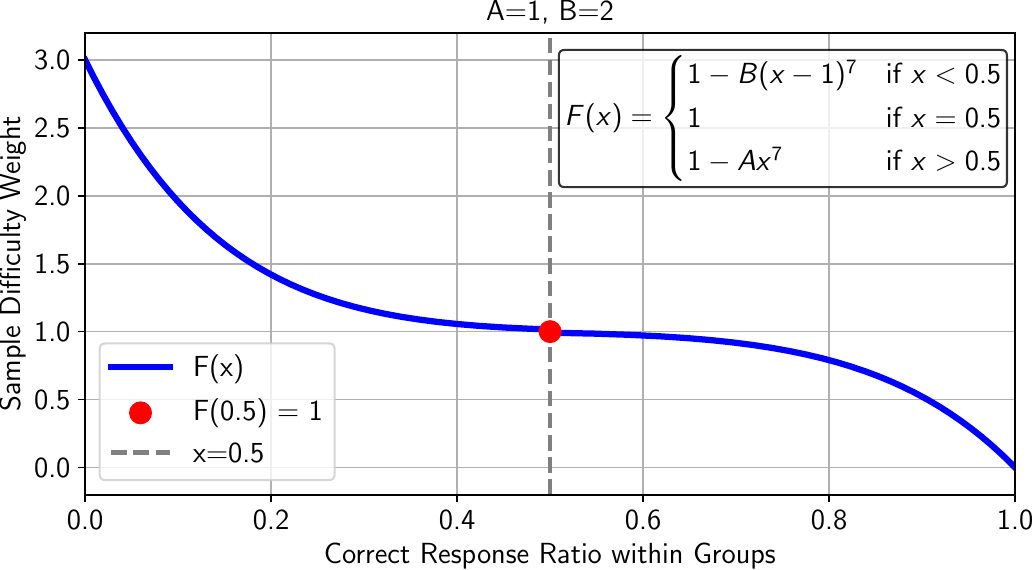}
\caption{Power Function for Sample Difficulty-Aware Reward Scaling}
\label{fig:math}
\end{figure}

To model this relationship, we employ a power function to map the correctness ratio (denoted as $r_c$) to the sample difficulty weight $w_s$, as shown in Figure~\ref{fig:math}. This function operationalizes the inverse relationship between correctness ratio and sample difficulty by assigning larger weights to harder samples and smaller weights to easier ones.
At the same time, the function is designed to preserve a key property of GRPO: when all responses receive identical rewards, the group advantages vanish. Specifically, when all responses are correct or all are incorrect, the group rewards are identical, resulting in zero advantage ($\hat{A}_i = 0$), regardless of the value of $w_s$. When approximately half of the responses are correct, the sample is of moderate difficulty and $w_s \approx 1$, resulting in minimal scaling. This design enables difficulty-aware reward scaling while maintaining compatibility with GRPO.

\subsection{{Objective Function of VULPO}}

Given the multidimensional reward design and difficulty-adaptive scaling, we obtain the group rewards $\mathbf{r}$ for responses $\{o_i\}_{i=1}^G$, along with the label difficulty weight $w_l$ and sample difficulty weight $w_s$. We compute the group advantages via mean-centering.
Unlike GRPO, we omit standard deviation normalization. Such normalization ignores difficulty variation across samples and may over-amplify easy samples (with consistently high rewards) while suppressing harder ones (with more diverse outcomes). In contrast, our difficulty-adaptive scaling explicitly accounts for sample difficulty, preventing excessive rewards for trivial cases while reinforcing learning on challenging ones.
The VULPO objective is defined as:
\begin{equation}\label{eq:vulpo} \begin{split} \mathcal{J}_{VULPO}(\theta) &= \mathbb{E}_{q \sim P(q), \{o_i\}_{i=1}^G \sim \pi_{\theta_{old}}(O|q)} \bigg[ \frac{1}{\sum_{i=1}^G|o_i|} \sum_{i=1}^G \sum_{t=1}^{|o_i|} \\ &\quad \Big(w_l\cdot w_s\cdot\hat{A}_{i,t} - \beta D_{KL} \big[ \pi_{\theta} || \pi_{ref} \big] \Big) \bigg], \end{split} \end{equation} $\text{where }\hat{A}_{i,t} = r_i-\text{mean}(\mathbf{r})$.

\noindent{\textbf{VULPO Algorithm.}} As summarized in Algorithm~\ref{alg:vulpo}, at each training step, a query batch $\mathcal{D}_n$ is sampled from VD dataset $\mathcal{D}$. For each input $q \in \mathcal{D}_n$, the current policy generates $G$ rollouts $\{o_i\}_{i=1}^G$. These rollouts are evaluated by the LLM-based reward model (Section~\ref{sec:reward_judge}) to obtain rewards $\{r_i\}_{i=1}^G$, from which group advantages $\{\hat{A}_{i,t}\}_{i=1}^G$ are computed via mean-centering.
The label difficulty weight $w_l$ is determined by the vulnerability label, while sample difficulty weight $w_s$ is computed based on the proportion of correct responses within the group. The policy is then updated by maximizing the VULPO objective (in Equation~\ref{eq:vulpo}) for $E$ epochs.

\begin{algorithm}[h]
  \caption{VULPO Training Procedure}
  \begin{algorithmic}[1]
  \footnotesize
    \Statex \hspace{-1.2em}\textbf{Input:} initial policy model $\pi_{\theta_{\text{init}}}$; VD Datasets $\mathcal{D}$
    \State policy model $\pi_\theta \leftarrow \pi_{\theta_{\text{init}}}$
    \For{epoch = 1, \dots, E}
       \State reference model $\pi_{ref} \leftarrow \pi_{\theta}$
      \For{step = 1, \dots, N}
      \State Sample a batch $\mathcal{D}_n$ from $\mathcal{D}$
      \State Update the old policy model $\pi_{\theta_{old}} \leftarrow \pi_{\theta}$ 
      \State Sample $G$ outputs $\{o_i\}_{i=1}^G \sim \pi_{\theta_{old}} (\cdot \mid q) $ for each question $q \in \mathcal{D}_n$
      \State $\{r_i\}_{i=1}^{G}\gets$\textbf{get\_rewards\_from\_judge}$(\{o_i\}_{i=1}^{G})$
      \State $w_l\gets$\textbf{compute\_label\_difficulty}$(q)$
      \State $w_s\gets$\textbf{compute\_sample\_difficulty}$(\{r_i\}_{i=1}^{G})$
      \State $\hat{A}_i \gets r_i - \text{mean}(\mathbf{r})$ // compute group advantage
      \State Update $\pi_{\theta}$ by maximizing the VULPO objective (Equation~\ref{eq:vulpo})
      \EndFor
    \EndFor 
    \Statex \hspace{-1.2em}\textbf{Output:} $\pi_\theta$
  \end{algorithmic}
  \label{alg:vulpo}
\end{algorithm}

\section{Experimental Setup}

\noindent\textbf{Datasets.}
We split ContextVul into training, validation, and test sets according to patch commit dates with a ratio of 8:1:1, resulting in 15,760, 1,970, and 1,970 samples, respectively. This time-ordered partitioning prevents future vulnerability information from appearing in the training set. To evaluate out-of-distribution (OOD) generalization, we additionally construct an unseen vulnerability benchmark consisting of newly disclosed CVEs whose patch commit dates are later than the knowledge cutoff dates of all evaluated LLMs. The details and evaluation results are presented in Section~\ref{sec:ood}.

\noindent\textbf{Models.}
We adopt the Qwen3 as the policy LLM in our experiments. In Section~\ref{sec:performance_comparison}, we present a comparative analysis of several general-purpose LLMs from the OpenAI, Gemini, Qwen, and DeepSeek series and specialized VD LLMs (i.e., R2VUL~\cite{37} and VulnLLM-R~\cite{45} under zero-shot settings, along with various post-training methods (i.e., R2VUL~\cite{37}, ReVD~\cite{36}, and MARCO~\cite{29}).

\noindent\textbf{Evaluation Method.}
During evaluation, we employ the same LLM-based judge protocol described in Section~\ref{sec:sft}. For vulnerable samples, correctly identifying the target vulnerability is considered a true positive (TP), while missing the target vulnerability or predicting unrelated vulnerabilities is treated as a false negative (FN). For patched samples, correctly recognizing the absence of the target vulnerability is considered a true negative (TN), whereas incorrectly predicting that the fixed vulnerability still exists is treated as a false positive (FP).

\noindent\textbf{Metrics.} As described in Section~\ref{sec:con}, our VD dataset is constructed from CVE patch commits, where each vulnerable code snippet has a corresponding patched version. An effective VD model should not only correctly identify the target vulnerability in the vulnerable code, but also recognize its absence in the patched version.
Accordingly, we propose Pairwise Pass@k (P-Pass@k) as the primary evaluation metric. For each vulnerability-patch query pair, we sample $k=8$ response pairs, where each pair consists of one response for the vulnerable code and one response for its patched counterpart. Let $p_{i,j} \in \{0,1\}$ denote whether the $j$-th response pair for the query pair $i$ is jointly correct, i.e., the model correctly identifies the target vulnerability in the vulnerable version and recognizes its absence in the patched version. Then, the P-Pass@k metrics are defined as: $\text{P-Pass@1} = \frac{1}{Nk}\sum\limits_{i=1}^{N}\sum\limits_{j=1}^{k} p_{i,j}$, $\text{P-Pass@8} = \frac{1}{N}\sum\limits_{i=1}^{N} [1 - \prod\limits_{j=1}^{k} (1 - p_{i,j}) ]$. We further introduce three pairwise consistency metrics, P-V, P-B, and P-R~\cite{1}, to characterize model behavior on vulnerability-patch pairs. We omit P-C since it is equivalent to P-Pass@1 in our setting. Specifically, P-V denotes cases where the model identifies the target vulnerability in both vulnerable and patched versions; P-B denotes cases where the model predicts the absence of the target vulnerability in both versions; and P-R denotes cases where the model produces reversed predictions with respect to the ground truth. In addition, we report the standard sample-level F1 score: $\text{F1} = \frac{2TP}{2TP + FP + FN}$.

\noindent\textbf{Implementation.}
We train all models using the Hugging Face TRL framework and perform inference with vLLM. For SFT, we use DeepSpeed ZeRO-3 on 4 A100 80GB GPUs. For RL training, we use 8 GPUs for policy optimization and an additional 4 GPUs for online rollout generation.
For SFT, we set the learning rate to $1\times10^{-5}$ with a linear scheduler, warmup ratio of $0.1$, weight decay of $0.01$, batch size of $32$, gradient accumulation steps of $8$, and training epochs of $3$. For RL training, we set the learning rate to $1\times10^{-6}$ with a linear scheduler, warmup ratio of $0.1$, batch size of $32$ (with each sample generating $8$ rollouts), gradient accumulation steps of $32$, and maximum gradient norm of $0.5$.

For optimization-based baselines, we follow the hyperparameter settings recommended in their original papers. For zero-shot prompting, we use the default or officially recommended sampling parameters for each evaluated LLM.
Like prior work, we set $\beta=0$ in Equation~\ref{eq:vulpo}, as we find it has negligible impact on performance while improving training efficiency. For reward evaluation during RL training and performance evaluation during testing, we use GPT-OSS-120B with high reasoning effort as the judge LLM to ensure accurate evaluation. In Section~\ref{rq:human}, we further investigate the consistency of evaluation results when using independent judge LLMs different from those used during reward evaluation.

In addition, we set maximum sequence length to 32,768 tokens, evenly divided between input and completion lengths (16,384 tokens each). This ensures that repository-level contextual information can be accommodated while reserving sufficient space for long-form vulnerability reasoning.

\section{Research Questions and Answers}

This section conducts experiments to analyze and discuss eight research questions (RQs).

\begin{itemize}[leftmargin=10pt]
\item\textbf{RQ1:} How does the cold-start stage affect RL performance in VULPO?

\item\textbf{RQ2:} How does the label difficulty weight affect RL performance in VULPO?

\item\textbf{RQ3:} How does VULPO perform compared to representative LLMs and VD baselines?

\item\textbf{RQ4:} Can VULPO generalize to unseen vulnerabilities during LLM pre-training?

\item\textbf{RQ5:} How does VULPO perform across different CWE categories compared to representative LLMs?

\item\textbf{RQ6:} To what extent do the multidimensional reward modules and difficulty-aware scaling strategies contribute to VULPO’s performance?

\item\textbf{RQ7:} How reliable are different judge LLMs for reward modeling and performance evaluation?

\item\textbf{RQ8:} How do different vulnerability evaluation granularities affect the assessment of LLM-based VD systems?

\end{itemize}

\subsection{RQ1: Impact of Cold-Start SFT}\label{sec:cold}

\begin{figure}[h]\centering
\includegraphics[width=0.9\columnwidth]{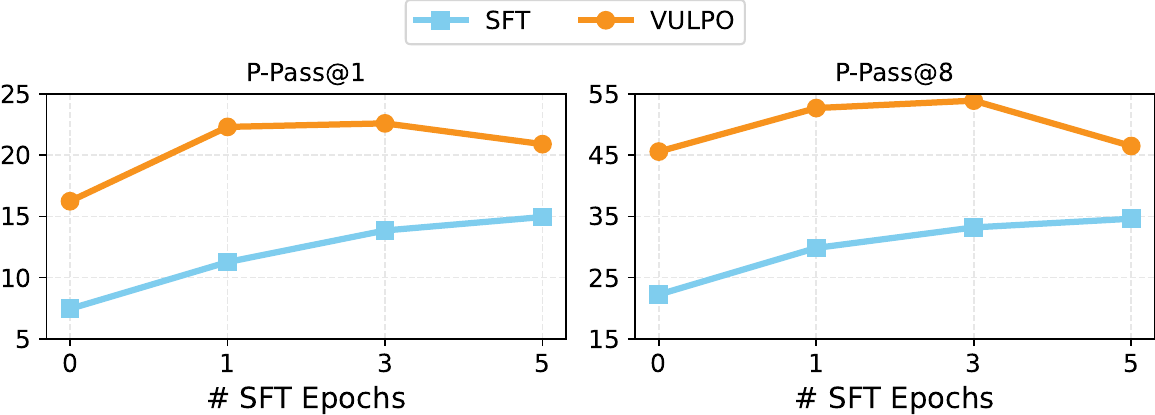}
\caption{Impact of Cold-Start SFT on VULPO Performance} \label{fig:RQ3}
\end{figure}

To investigate the role of the cold-start stage in subsequent RL training, we compare VULPO performance under different numbers of SFT epochs prior to RL. 

From Figure~\ref{fig:RQ3}, we can find that directly applying RL to the base Qwen3-4B model without task-specific cold-start SFT leads to substantially worse performance compared to models initialized with SFT. This phenomenon arises because the VD task is highly challenging for small-scale LLMs at initialization. Without prior exposure to vulnerability reasoning patterns, the model struggles to generate correct rollouts during on-policy RL training, resulting in sparse and low-quality reward signals that hinder effective model exploration.

Notably, introducing a lightweight cold-start stage substantially improves VULPO performance. With 1-3 epochs of SFT, the model achieves significantly higher P-Pass@1 and P-Pass@8 after RL training. This suggests that the cold-start stage effectively bootstraps the model’s ability to generate informative vulnerability reasoning trajectories, thereby increasing the probability of obtaining meaningful rewards during RL.
However, increasing the number of SFT epochs beyond a moderate level leads to degraded RL performance, particularly in P-Pass@8. We hypothesize that excessive SFT causes the model to overfit to fixed reasoning patterns in the curated reasoning dataset, reducing rollout diversity and limiting model exploration. As a result, the policy becomes less capable of discovering improved reasoning strategies through on-policy optimization.

Overall, these results demonstrate that a lightweight cold-start stage is critical for balancing initial rollout correctness and exploration ability in RL-based VD. Moderate SFT initialization provides sufficiently informative starting policies while preserving the diversity necessary for effective RL exploration, ultimately leading to the best downstream performance.

\subsection{RQ2: Impact of label difficulty weight}\label{sec:labeldiff}

\begin{figure}[h]\centering
\includegraphics[width=0.9\columnwidth]{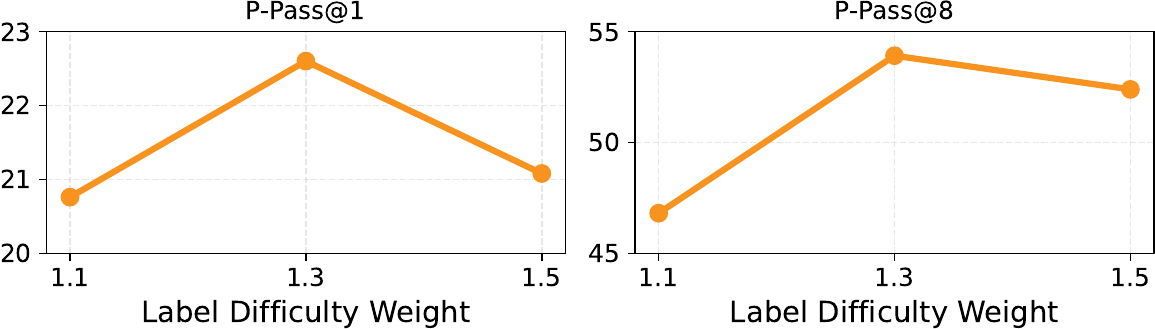}
\caption{Impact of Label Difficulty Weight on VULPO Performance} \label{fig:RQ4}
\end{figure}

In VD, correctly identifying the target CVE in vulnerable code is substantially more difficult than recognizing its absence in patched code. As discussed in Section~\ref{sec:mtv}, this asymmetry in prediction difficulty can bias RL training toward easier negative predictions, potentially leading to reward hacking and conservative detection behavior.
To investigate the impact of label-difficulty scaling, we vary the label difficulty weight and evaluate the performance of VULPO. As shown in Figure~\ref{fig:RQ4}, when a smaller label difficulty weight is used, the model achieves relatively low P-Pass@8, indicating limited exploration and conservative vulnerability detection behavior. In this setting, the reward gap between true positive and true negative predictions is insufficient to compensate for the higher difficulty of identifying vulnerabilities, causing the model to favor easier negative predictions during RL training.

Increasing the label difficulty weight substantially improves both P-Pass@1 and P-Pass@8. This suggests that assigning larger rewards to correctly identified vulnerable samples effectively counteracts the inherent asymmetry between positive and negative predictions, encouraging the model to explore more informative vulnerability reasoning trajectories.
However, excessively large label difficulty weights lead to performance degradation. We hypothesize that overly amplified rewards for vulnerable samples bias the model toward a subset of high-reward positive patterns, reducing generalization across diverse vulnerability types and harming overall detection performance.

In summary, moderate label-difficulty scaling is critical for stable and effective RL training in VD. Properly balancing the reward gap between true positive and true negative predictions mitigates reward hacking while maintaining strong vulnerability sensitivity and generalization ability.

\subsection{RQ3: Performance Comparison on ContextVul}\label{sec:performance_comparison}

\begin{table}[h]
\caption{Performance Comparison on ContextVul}
\label{tab:RQ1}
\centering
\renewcommand{\arraystretch}{1}
\begin{threeparttable}
\resizebox{\columnwidth}{!}{
\begin{tabular}{ccccccc}
\toprule
                                 & \textbf{P-Pass@1} & \textbf{P-Pass@8} & \textbf{F1} & \textbf{P-B$\downarrow$} & \textbf{P-V$\downarrow$} & \textbf{P-R$\downarrow$} \\ \hline
\textbf{Zero-Shot Prompting}     &                 &                 &             &                          &                          &                          \\
Qwen3-4B                         & 7.45           & 22.23           & 15.20       & 87.61                    & 1.18                     & 3.76                     \\
R2VUL-7B~\cite{37}               & 3.78           & 16.24           & 14.68       & 81.33                    & 5.32                     & 9.57                     \\
VulnLLM-R-7B~\cite{45}           & 5.08           & 21.73           & 12.18       & 89.61                    & 1.75                     & 3.57                     \\
DeepSeek-R1-0528-Qwen3-8B        & 11.78          & 32.59           & 22.48       & 82.82                    & 1.57                     & 3.83                     \\
GPT-OSS-20B                      & 12.98          & 26.09           & 26.18       & 81.62                    & 2.89                     & 2.50                     \\
GPT-OSS-120B                     & \underline{22.16} & 40.10        & \underline{39.58} & 71.03              & 4.20                     & 2.61                     \\
o4-mini                          & 15.98          & 31.78           & 29.94       & 80.74                    & 2.21                     & 1.08                     \\
Qwen3-235B-A22B                  & 16.26          & 35.43           & 29.86       & 78.55                    & 2.21                     & 2.98                     \\
DeepSeek-R1-0528                 & 21.55          & \underline{42.03} & 37.32     & 69.37                    & 3.48                     & 5.61                     \\
DeepSeek-V3.1                    & \underline{22.50} & \underline{42.13} & \underline{39.64} & 71.95         & 3.59                     & 1.95                     \\
gemini-2.5-flash                 & 19.66          & 39.29           & 35.85       & 75.18                    & 3.31                     & 1.85                     \\ \midrule
\textbf{Off-Policy Optimization} &                 &                 &             &                          &                          &                          \\
Qwen3-4B-SFT                     & 14.95          & 34.62           & 28.26       & 79.70                    & 2.39                     & 2.97                     \\
Qwen3-4B-R2VUL~\cite{37}         & 5.36           & 25.99           & 12.26       & 84.87                    & 1.81                     & 7.96                     \\
Qwen3-4B-ReVD~\cite{36}          & 4.63           & 23.25           & 11.28       & 78.19                    & 2.37                     & 14.81                    \\
Qwen3-4B-DPO                     & 16.94          & 40.30           & 31.28       & 76.05                    & 2.89                     & 4.11                     \\
Qwen3-4B-ORPO                    & 15.13          & 36.24           & 28.79       & 77.42                    & 2.94                     & 4.51                     \\ \midrule
\textbf{On-Policy RL}            &                 &                 &             &                          &                          &                          \\
Qwen3-4B-MARCO~\cite{29}         & 11.52          & 27.51           & 22.24       & 85.51                    & 1.36                     & 1.61                     \\
VULPO-4B (Ours)                  & \textbf{22.60} & \textbf{53.91} & \textbf{40.09} & 65.77                 & 5.38                     & 6.24                     \\ \bottomrule
\multicolumn{7}{l}{\footnotesize \textbf{Bold} indicates the best result, while \underline{underline} denotes the second-best result.}
\end{tabular}
}
\end{threeparttable}
\end{table}

In this section, we compare VULPO against representative LLMs and VD baselines across three paradigms: zero-shot prompting, off-policy optimization, and on-policy RL. To ensure fair comparison, all optimization-based methods are trained using the same backbone model, Qwen3-4B. We additionally include the original R2VUL-7B checkpoint trained on Qwen2.5-7B-Instruct. Since ReVD and MARCO have not publicly released model weights, and the publicly available ReVD implementation is insufficient for exact reproduction, we contact the authors and make our best effort to reproduce their methods after receiving no response.

Table~\ref{tab:RQ1} shows VULPO-4B achieves  best overall performance across all evaluation metrics, obtaining a P-Pass@1 of 22.60\%, P-Pass@8 of 53.91\%, and F1 score of 40.09\%. Notably, despite using only a 4B backbone model, VULPO surpasses substantially larger reasoning LLMs, including DeepSeek-V3.1, DeepSeek-R1-0528, and GPT-OSS-120B.
Compared with the base Qwen3-4B model, VULPO relatively improves P-Pass@1 by 203.4\% and P-Pass@8 by 142.5\%. Even compared with Qwen3-4B-SFT, VULPO still achieves relative improvements of 51.2\% in P-Pass@1 and 55.7\% in P-Pass@8, demonstrating the effectiveness of the proposed process-centric RL optimization framework.

However, several off-policy optimization methods, including R2VUL and ReVD, perform even worse than the original Qwen3-4B baseline. We attribute this degradation to two major limitations. First, these methods introduce ground-truth vulnerability information when prompting teacher LLMs during reasoning data construction, leading to information leakage and unrealistic reasoning supervision. Consequently, the distilled student models tend to produce hallucinated vulnerability analyses during inference. Second, these approaches rely primarily on binary-detection correctness or coarse CWE-level filtering to select reasoning traces, resulting in noisy and low-quality supervision signals that fail to capture the quality of the underlying reasoning process.

\subsection{RQ4: Out-of-Distribution Generalization on Unseen CVEs}\label{sec:ood}

\begin{figure}[h]\centering
\includegraphics[width=\columnwidth]{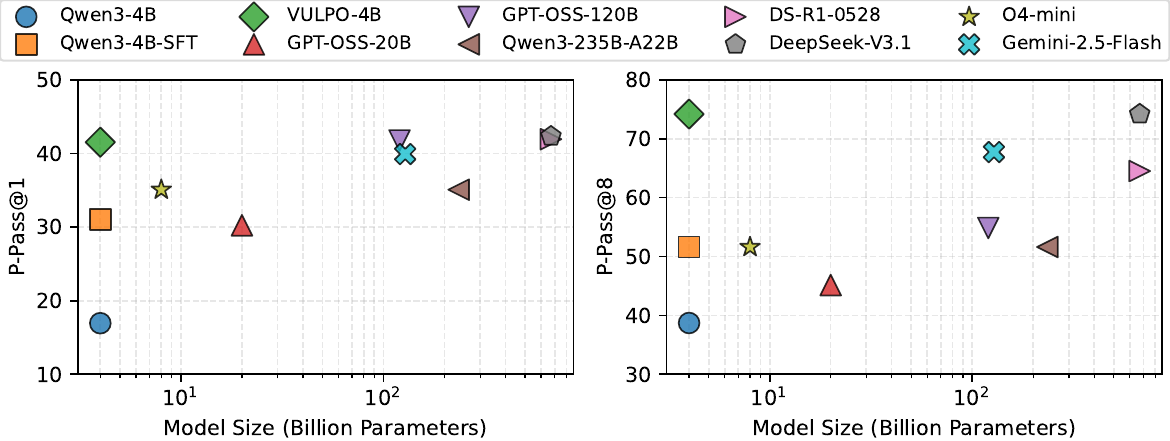}
\caption{Out-of-Distribution Generalization on Unseen CVEs}
\label{fig:ood}
\end{figure}

A common concern in LLM-based VD research is the potential overlap between LLM pre-training corpora and publicly disclosed vulnerability data. Since modern LLMs are continuously updated while newly disclosed CVEs remain relatively limited, completely eliminating potential data leakage during evaluation is inherently challenging.

Recall Table~\ref{tab:RQ1} demonstrates VULPO substantially improves VD capability of the underlying backbone model on the same evaluation set, we further investigate whether the observed gains generalize to truly unseen vulnerabilities.
Toward this, we construct an OOD evaluation set consisting of CVEs whose patch commit dates are later than the knowledge cutoff dates of all evaluated LLMs. Following prior work~\cite{1}, we restrict the collected vulnerability patches to commits that modify only a single function within a single file, ensuring accurate vulnerability labeling and clean vulnerable-patched pairs. After data collection and filtering, the final OOD benchmark contains 31 vulnerability-patch pairs. We further follow the same context extraction pipeline described in Section~\ref{sec:con} to construct repository-level contextual information for evaluation.

Figure~\ref{fig:ood} presents OOD generalization results. Overall, the relative performance trends remain highly consistent with those observed on the in-distribution benchmark in Table~\ref{tab:RQ1}, suggesting that the superior performance of VULPO cannot be simply attributed to potential pre-training data leakage.

Notably, VULPO achieves stronger P-Pass@1 and P-Pass@8 performance than Qwen3-235B-A22B, Gemini-2.5-Flash, and DeepSeek-R1-0528, while remaining competitive with GPT-OSS-120B and DeepSeek-V3.1. These results demonstrate that the proposed process-centric RL framework generalizes effectively beyond memorized knowledge and enables robust vulnerability reasoning on unseen CVEs.

\subsection{RQ5: Recall Comparison Across Different CWE Types}

\begin{figure}[h]
\centering
\includegraphics[width=0.45\textwidth]{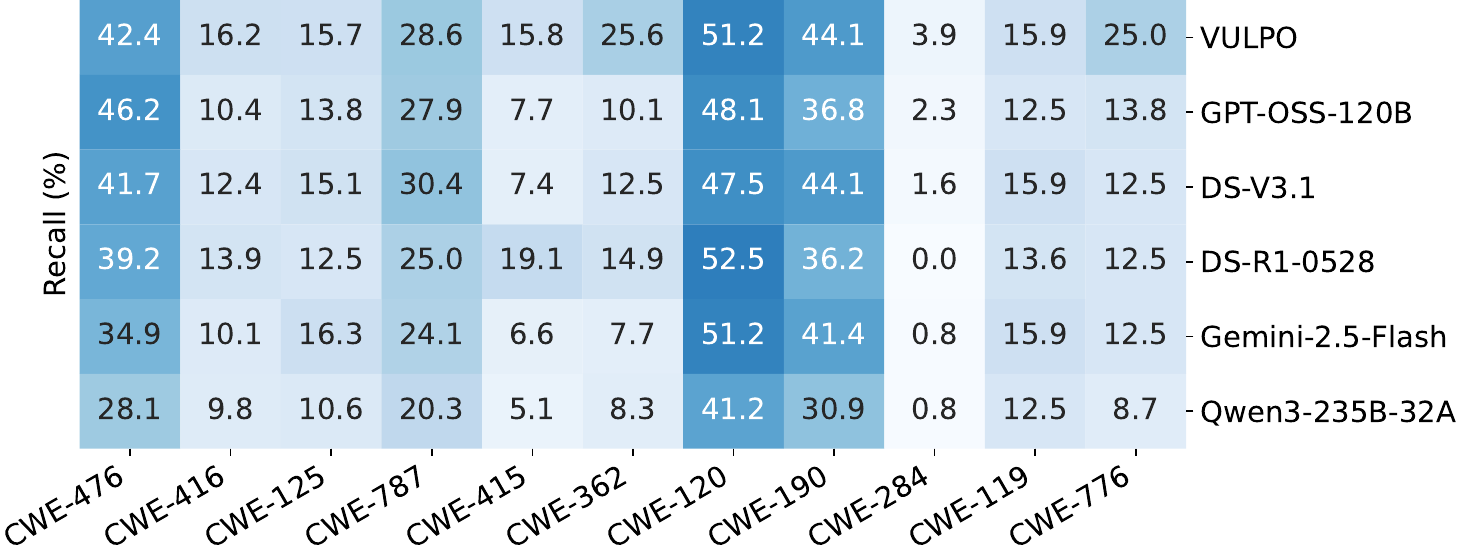}
\caption{Recall Comparison Across Different CWE Types}
\label{fig:cwe}
\end{figure}

To better understand the strengths and limitations of different LLMs, we further analyze model performance across representative CWE categories. Figure~\ref{fig:cwe} reports the recall of different models on major vulnerability types in ContextVul.
Overall, all evaluated LLMs achieve relatively strong performance on several classical vulnerability categories, particularly CWE-120 (Buffer Copy without Checking Size of Input) and CWE-190 (Integer Overflow or Wraparound). These vulnerability types are typically associated with highly recognizable code patterns and high-risk primitives, such as unsafe buffer operations or arithmetic on user-controlled integers.

In contrast, VULPO indicates clear advantages on reasoning-intensive vulnerability categories, especially memory-management and concurrency-related vulnerabilities. In particular, VULPO achieves the best performance on CWE-416 (Use After Free), CWE-415 (Double Free), and CWE-362 (Race Condition), substantially outperforming other frontier LLMs.
Unlike classical buffer overflows, these vulnerabilities require reasoning over long-range program semantics, including pointer lifetimes, resource ownership, synchronization states, and inter-procedural control flows. For example, detecting use-after-free vulnerabilities often requires tracking whether freed objects remain reachable across multiple execution paths, while race-condition detection depends on understanding concurrent access patterns and synchronization semantics. The superior performance of VULPO on these categories suggests that process-centric RL training effectively improves the model’s ability to perform context-aware vulnerability reasoning rather than relying solely on superficial API-level patterns.

Finally, all evaluated models perform poorly on CWE-284 (Improper Access Control), revealing an important limitation of current LLM-based VD systems. Unlike memory corruption vulnerabilities, access control issues are often omission-based vulnerabilities that arise from missing authorization or validation logic rather than explicitly unsafe operations. Detecting such vulnerabilities requires inferring implicit security invariants and intended protection semantics, which remains highly challenging for current LLMs.

\subsection{RQ6: Ablation Study}\label{sec:ablation_study}

\begin{figure}[h]\centering
\includegraphics[width=0.8\columnwidth]{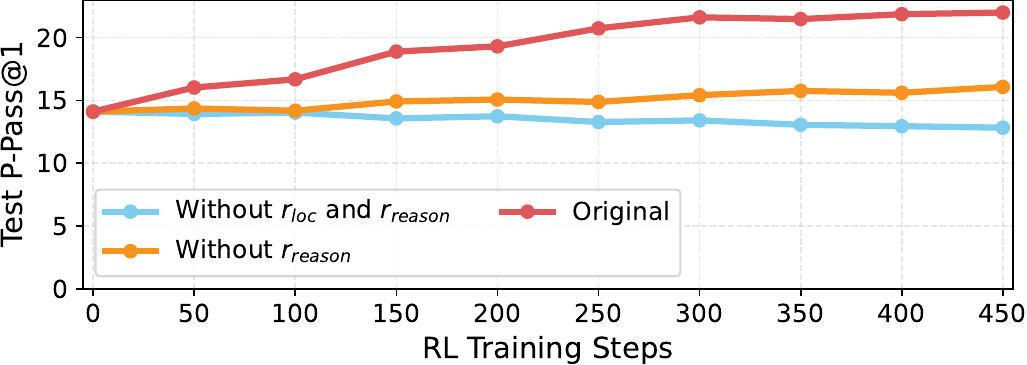}
\caption{Ablation Study on Process-Aware Multidimensional Rewards}
\label{fig:ablation_reward}
\end{figure}


\noindent\textbf{Ablation on Process-Aware Multidimensional Rewards.} To investigate the contribution of different reward dimensions, we conduct ablation experiments by progressively removing process-level reward components during RL training. Specifically, we compare the full VULPO reward design against two degraded variants: (i) removing the causal reasoning reward while retaining outcome correctness and localization rewards, and (ii) removing both causal reasoning and localization rewards, leaving only outcome correctness supervision.

Figure~\ref{fig:ablation_reward} presents the RL training dynamics under different reward configurations. When both localization and reasoning rewards are removed, the model exhibits the worst overall performance and unstable training behavior. This result confirms that outcome-only supervision provides insufficient guidance for context-aware vulnerability reasoning, causing RL optimization to rely primarily on coarse prediction correctness.

Introducing localization rewards improves training stability and overall performance to some extent, indicating that grounding vulnerability predictions in relevant code regions provides more informative supervision signals than binary outcome correctness alone. However, the largest improvement is achieved after incorporating causal reasoning rewards. With full process-aware rewards, the model consistently attains the highest P-Pass@1 throughout RL training.

These results verify that fine-grained process-level supervision is critical for effective RL-based VD. In particular, the causal reasoning reward substantially improves the model’s ability to learn structured vulnerability analysis beyond superficial pattern matching, validating the effectiveness of the proposed process-aware multidimensional reward design.

\begin{table}[h]
\caption{Ablation Study on Difficulty-Aware Reward Scaling}
\label{tab:RQ5}
\centering
\begin{threeparttable}
\resizebox{\columnwidth}{!}{
\renewcommand{\arraystretch}{1}
\begin{tabular}{cccccc}
\toprule
\textbf{Module} & \textbf{\begin{tabular}[c]{@{}c@{}}Label Difficulty\\ Reward Scaling\end{tabular}} & \textbf{\begin{tabular}[c]{@{}c@{}}Sample Difficulty\\ Reward Scaling\end{tabular}} & \textbf{P-Pass@1} & \textbf{P-Pass@8} & \textbf{F1} \\ \midrule
\scalebox{1.5}{\ding{172}} &  &  & 19.91 & 44.87 & 36.67 \\
\scalebox{1.5}{\ding{173}} & \scalebox{1.5}{\ding{51}} &  & 20.58 & 47.21 & 38.52 \\
\scalebox{1.5}{\ding{174}} &  & \scalebox{1.5}{\ding{51}} & 20.13 & 46.19 & 37.33 \\
VULPO & \scalebox{1.5}{\ding{51}} & \scalebox{1.5}{\ding{51}} & \textbf{22.60} & \textbf{53.91} & \textbf{40.09} \\ \bottomrule
\end{tabular}
}
\end{threeparttable}
\end{table}

\noindent\textbf{Ablation on Difficulty-Aware Reward Scaling.}
Table~\ref{tab:RQ5} presents ablation results for proposed difficulty-aware reward scaling strategies. Specifically, we evaluate the effects of label-level scaling and sample-level scaling individually and jointly.

Without any difficulty-aware scaling, the RL-trained model achieves relatively limited performance, especially on P-Pass@8, showing insufficient exploration during RL training. Introducing label difficulty scaling yields largest single-module improvement across all metrics. This validates our motivation that vulnerable samples are inherently more difficult than non-vulnerable ones, and amplifying rewards for correctly identified vulnerable samples effectively mitigates reward hacking caused by prediction difficulty asymmetry.

Sample difficulty scaling also consistently improves model performance. By assigning larger rewards to difficult samples with lower rollout correctness ratios, the model is encouraged to explore more challenging reasoning trajectories, alleviating conservative detection behavior during RL optimization.

Finally, combining both label-level and sample-level scaling yields the best overall performance. Compared with using only label difficulty scaling, the full VULPO framework further improves P-Pass@1 from 20.58\% to 22.60\% and P-Pass@8 from 47.21\% to 53.91\%, demonstrating that the two scaling strategies play complementary roles in improving RL exploration and vulnerability reasoning capability.

\subsection{RQ7: Impact of Judge LLMs on Evaluation Reliability}\label{rq:human}

\begin{table}[h]
\caption{Evaluation Bias of Different Judge LLMs}
\label{tab:judge_bias}
\centering
\resizebox{0.85\columnwidth}{!}{
\renewcommand{\arraystretch}{1.0}
\begin{tabular}{ccc}
\toprule
\textbf{Judge LLM} & \textbf{\# Correct Judgments} & \textbf{\# Incorrect Judgments} \\ \midrule
GPT-4.1 mini        & 341                      & 59                          \\
GPT-OSS-20B         & 395                      & 5                           \\
GPT-OSS-120B        & \textbf{399}             & \textbf{1}                 \\ \bottomrule
\end{tabular}
}
\end{table}

Since our framework relies heavily on LLM-based reward modeling and evaluation, the reliability of judge LLMs is critical for ensuring trustworthy RL optimization and accurate model assessment. To investigate the impact of judge selection, we benchmark three candidate judge LLMs: GPT-4.1 mini, GPT-OSS-20B, and GPT-OSS-120B. For all judge models, we enable the highest available reasoning effort settings when applicable.
To quantify evaluation accuracy, we randomly sample 400 model completions generated by Qwen3-4B and manually verify the correctness of judge predictions. Table~\ref{tab:judge_bias} reports the corresponding evaluation results.

Among the evaluated judges, GPT-4.1 mini exhibits substantially lower reliability, incorrectly judging 59 out of 400 samples, corresponding to an error rate of 14.75\%. Our analysis reveals that most errors occur when the predicted vulnerability category is partially aligned with the ground-truth CVE, but the underlying reasoning process is flawed. In such cases, GPT-4.1 mini often focuses primarily on surface-level vulnerability keywords while failing to rigorously verify the correctness of the root-cause analysis.

In contrast, GPT-OSS-20B and GPT-OSS-120B both achieve over 99\% judgment accuracy. The few remaining errors are primarily caused by unfaithful reasoning, where the judge incorrectly claims that the model fails to identify vulnerabilities despite explicit vulnerability predictions appearing in the response. We additionally observe that GPT-OSS-20B occasionally produces repetitive or unstable reasoning outputs during evaluation. Therefore, we adopt GPT-OSS-120B as the default judge LLM throughout this work to ensure stable and accurate reward assessment during RL training and reliable performance evaluation during testing.

\begin{table}[h]
\caption{Performance Evaluation Using an Independent Judge LLM}
\label{tab:judge_independence}
\centering
\resizebox{\columnwidth}{!}{
\renewcommand{\arraystretch}{1.0}
\begin{tabular}{ccccc}
\toprule
\textbf{Judge LLM} & \textbf{P-Pass@1} & \textbf{P-Pass@8} & \textbf{F1} & \textbf{Cohen's Kappa} \\ \midrule
GPT-OSS-120B & 22.60 & 53.91 & 40.09 & \multirow{2}{*}{0.8115} \\
Qwen3.5-122B-A10B & 22.32 & 54.01 & 39.04 &  \\ \bottomrule
\end{tabular}
}
\end{table}

\noindent\textbf{Mitigating Judge Overfitting via Independent Evaluation.}
A potential concern in LLM-based RL frameworks is judge overfitting, where the policy model may implicitly optimize toward the specific preferences or biases of the reward judge rather than genuinely improving reasoning capability. To investigate this issue, we decouple the reward and evaluation judges by using GPT-OSS-120B exclusively for reward computation during RL training, while employing Qwen3.5-122B-A10B as an independent judge for final evaluation.

Table~\ref{tab:judge_independence} shows that the evaluation results produced by the two judge LLMs remain highly consistent, achieving a Cohen’s Kappa coefficient of 0.8115. The overall performance metrics remain nearly unchanged under the independent judge setting. These results suggest that the performance gains achieved by VULPO are not artifacts of overfitting to a specific reward judge, but instead reflect genuine improvements in context-aware vulnerability reasoning capability.

\subsection{RQ8: Evaluation Bias Across Hierarchical Correctness}\label{sec:sankey}


\begin{figure}[h]\centering
\includegraphics[width=0.38\textwidth]{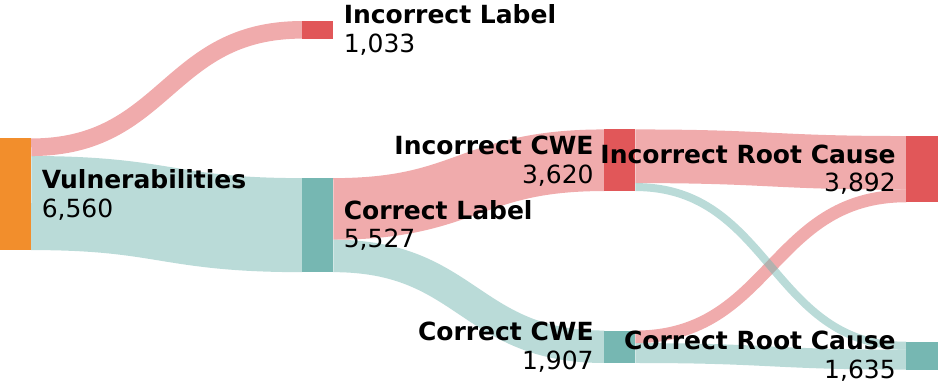}
\caption{Hierarchical Correctness Evaluation of Vulnerability Predictions}
\label{fig:sankey}
\end{figure}

Existing VD evaluation protocols predominantly rely on binary vulnerability labels or CWE-level classification accuracy. However, such coarse-grained evaluation criteria may substantially overestimate a model’s true vulnerability reasoning capability. To investigate this, we conduct a hierarchical correctness analysis over 6,560 completions generated by DeepSeek-R1-0528 on the ContextVul test set.
To ensure fair comparison across different correctness granularities, we exclude samples without ground-truth CWE annotations. We then evaluate model outputs under three progressively stricter correctness criteria: (i) binary vulnerability detection, (ii) CWE type classification, and (iii) root-cause identification.

Figure~\ref{fig:sankey} illustrates how predictions progressively fail under increasingly fine-grained evaluation standards. In binary detection, the model correctly identifies 5,527 vulnerabilities as vulnerable. However, among these successful detections, 65.5\% are assigned incorrect CWE types. This result suggests that binary-label evaluation can substantially inflate reported VD performance, since correctly predicting the existence of a vulnerability does not necessarily imply understanding its actual vulnerability category or underlying cause.

Even under CWE-level evaluation, substantial discrepancies remain between CWE correctness and genuine vulnerability understanding. Among 1,907 predictions with correct CWE classifications, 14.3\% still contain incorrect or unrelated root-cause explanations, which suggests that predicting the correct CWE type alone is insufficient for assessing whether a model truly understands the vulnerability reasoning process.

Besides, CWE-level evaluation itself introduces additional bias due to the hierarchical and overlapping nature of CWE taxonomies. In many cases, models predict CWE categories that differ from the annotated labels but remain semantically consistent with the true vulnerability root cause. We observe that among the 3,620 predictions with incorrect CWE labels, 7.5\% still correctly identify the ground-truth root cause. Under strict exact-match CWE evaluation, such predictions are incorrectly treated as entirely wrong despite containing valid vulnerability reasoning.

Collectively, these findings demonstrate that binary-label and CWE-level evaluations systematically overestimate or distort the true capability of LLM-based VD systems. Accurate assessment of vulnerability understanding requires root-cause-level evaluation, which more faithfully reflects whether the model genuinely understands the vulnerability rather than merely recognizing superficial patterns or coarse-grained labels.
Importantly, this hierarchy of correctness also highlights the necessity in process-centric RL. Without fine-grained reward signals grounded in causal reasoning quality, RL training may incorrectly reward flawed reasoning trajectories that happen to produce correct binary predictions or CWE labels, ultimately limiting the reliability of VD models.
\section{Related Work}

\noindent\textbf{Prompt Engineering.}
Early LLM-based VD studies mainly use carefully designed prompts to improve vulnerability reasoning. Some methods guide LLMs with step-by-step instructions that imitate human security experts~\cite{4,5,6}, while others incorporate structural and semantic code information to help detect vulnerability patterns~\cite{4,32}. To overcome the limitations of a single LLM, several works adopt multi-agent frameworks where different agents collaboratively perform vulnerability analysis~\cite{8,33,34,35}. However, these approaches still heavily depend on the reasoning capability of the underlying base LLM, which is usually not specifically trained for vulnerability analysis.

\noindent\textbf{Supervised Fine-Tuning Methods.}
Some studies adapt LLMs for VD by adding classification heads for vulnerability prediction~\cite{15,20,21}. Other works use stronger teacher LLMs to generate CoT reasoning data and then fine-tune lightweight student LLMs through SFT~\cite{39,45}. Although these methods improve VD performance, classification-based approaches often lack interpretability, while SFT-based models may overfit to fixed vulnerability patterns and struggle to generalize to unseen vulnerabilities.

\noindent\textbf{Off-Policy Preference Optimization.}
Recent advances in RL from human feedback~\cite{38} and preference optimization methods such as DPO~\cite{22} and ORPO~\cite{23,40} have motivated several studies to optimize LLMs for VD using preference-based supervision. For example, ReVD~\cite{36} constructs preference pairs by prompting teacher LLMs to generate forward and backward reasoning trajectories from vulnerable and patched code. The student LLM is then optimized through curriculum-based identity preference optimization (IPO). Similarly, R2Vul~\cite{37} instructs a teacher LLM to generate valid reasoning traces using ground-truth vulnerability labels while constructing rejected responses through label flipping, followed by ORPO-based optimization of the student model.
In particular, these methods rely on vulnerability-related metadata (e.g., CVE descriptions or commit messages) to guide teacher LLMs during reasoning generation (i.e., rationalization). Consequently, the generated reasoning traces may implicitly depend on information unavailable during inference, potentially introducing a mismatch between training-time supervision and deployment-time inputs. This mismatch can limit the student model’s ability to generalize robust vulnerability reasoning solely from code and contextual information. Furthermore, these approaches are fundamentally based on static optimization. Such training paradigms restrict the model’s ability to dynamically explore diverse reasoning trajectories and adapt its policy according to feedback from its own online rollouts.

\noindent\textbf{On-Policy Optimization.}
Although on-policy RL methods such as GRPO have recently demonstrated strong effectiveness for improving reasoning capabilities in general LLM tasks, their applications to VD remain largely underexplored. MARCO~\cite{29} attempts to improve CoT reasoning for VD using GRPO-based optimization. However, its reward design primarily relies on coarse-grained binary correctness signals and does not explicitly model the reasoning process underlying vulnerability analysis. As a result, high reward values do not necessarily translate into reliable VD capability.
\section{Discussion}

\noindent\textbf{Evaluation Scope.}
Due to computational constraints, we train VULPO only on a 4B-scale backbone (Qwen3-4B). Despite its small scale, VULPO achieves performance competitive with much larger frontier LLMs, demonstrating the effectiveness of the proposed framework. Extending VULPO to larger models is left for future work.
Our experiments focus on C/C++ VD, but the framework is inherently programming-language agnostic. Building high-quality context-aware VD datasets for more programming languages remains an open challenge.

\noindent\textbf{Benefits for Downstream Cybersecurity Tasks.}
Beyond VD, VULPO’s context-aware reasoning capability may support downstream tasks such as PoC generation and vulnerability patching, highlighting its potential for broader cybersecurity tasks. We leave these directions to future work.

\section{Conclusion}

This work addressed context-aware vulnerability detection (VD) by modeling inter-procedural interactions often overlooked by function-level analyses and prior learning-based approaches. First, we constructed ContextVul, a dataset enriched with repository-level contextual information and vulnerability reasoning traces. Building upon this dataset, we proposed a two-stage post-training framework centered on vulnerability-adaptive policy optimization (VULPO) for context-aware vulnerability reasoning. Extensive experiments and ablation studies validated the effectiveness of the proposed framework.

\section*{Ethical Considerations}
Our work aims to train specialized LLMs for vulnerability detection tasks. All experiments are conducted using publicly available data, models, and source code. Since no new vulnerabilities are generated or identified in this work, no ethical issues will arise.

\bibliographystyle{IEEEtran}
\bibliography{sample-base}

@inproceedings{1,
  author       = {Yangruibo Ding and
                  Yanjun Fu and
                  Omniyyah Ibrahim and
                  Chawin Sitawarin and
                  Xinyun Chen and
                  Basel Alomair and
                  David A. Wagner and
                  Baishakhi Ray and
                  Yizheng Chen},
  title        = {Vulnerability Detection with Code Language Models: How Far are We?},
  booktitle    = {47th {IEEE/ACM} International Conference on Software Engineering,
                  {ICSE} 2025, Ottawa, ON, Canada, April 26 - May 6, 2025},
  pages        = {1729--1741},
  publisher    = {{IEEE}},
  year         = {2025}
}

@inproceedings{2,
  author       = {Xinchen Wang and
                  Ruida Hu and
                  Cuiyun Gao and
                  Xin{-}Cheng Wen and
                  Yujia Chen and
                  Qing Liao},
  title        = {ReposVul: {A} Repository-Level High-Quality Vulnerability Dataset},
  booktitle    = {Proceedings of the 2024 {IEEE/ACM} 46th International Conference on
                  Software Engineering: Companion Proceedings, {ICSE} Companion 2024,
                  Lisbon, Portugal, April 14-20, 2024},
  pages        = {472--483},
  publisher    = {{ACM}},
  year         = {2024}
}

@inproceedings{3,
  author       = {Ziyang Li and
                  Saikat Dutta and
                  Mayur Naik},
  title        = {{IRIS:} LLM-Assisted Static Analysis for Detecting Security Vulnerabilities},
  booktitle    = {The Thirteenth International Conference on Learning Representations,
                  {ICLR} 2025, Singapore, April 24-28, 2025},
  publisher    = {OpenReview.net},
  year         = {2025}
}

@inproceedings{4,
  author       = {Yansong Li and
                  Paula Branco and
                  Alexander M. Hoole and
                  Manish Marwah and
                  Hari Manassery Koduvely and
                  Guy{-}Vincent Jourdan and
                  Stephan Jou},
  editor       = {Marina Blanton and
                  William Enck and
                  Cristina Nita{-}Rotaru},
  title        = {SV-TrustEval-C: Evaluating Structure and Semantic Reasoning in Large
                  Language Models for Source Code Vulnerability Analysis},
  booktitle    = {{IEEE} Symposium on Security and Privacy, {SP} 2025, San Francisco,
                  CA, USA, May 12-15, 2025},
  pages        = {3014--3032},
  publisher    = {{IEEE}},
  year         = {2025}
}

@inproceedings{5,
  author       = {Saad Ullah and
                  Mingji Han and
                  Saurabh Pujar and
                  Hammond Pearce and
                  Ayse K. Coskun and
                  Gianluca Stringhini},
  title        = {LLMs Cannot Reliably Identify and Reason About Security Vulnerabilities
                  (Yet?): {A} Comprehensive Evaluation, Framework, and Benchmarks},
  booktitle    = {{IEEE} Symposium on Security and Privacy, {SP} 2024, San Francisco,
                  CA, USA, May 19-23, 2024},
  pages        = {862--880},
  publisher    = {{IEEE}},
  year         = {2024}
}

@inproceedings{6,
  author       = {Yu Nong and
                  Haoran Yang and
                  Long Cheng and
                  Hongxin Hu and
                  Haipeng Cai},
  editor       = {Lujo Bauer and
                  Giancarlo Pellegrino},
  title        = {{APPATCH:} Automated Adaptive Prompting Large Language Models for
                  Real-World Software Vulnerability Patching},
  booktitle    = {34th {USENIX} Security Symposium, {USENIX} Security 2025, Seattle,
                  WA, USA, August 13-15, 2025},
  pages        = {4481--4500},
  publisher    = {{USENIX} Association},
  year         = {2025}
}

@inproceedings{7,
  author       = {Ahmed Lekssays and
                  Hamza Mouhcine and
                  Khang Tran and
                  Ting Yu and
                  Issa Khalil},
  editor       = {Lujo Bauer and
                  Giancarlo Pellegrino},
  title        = {LLMxCPG: Context-Aware Vulnerability Detection Through Code Property
                  Graph-Guided Large Language Models},
  booktitle    = {34th {USENIX} Security Symposium, {USENIX} Security 2025, Seattle,
                  WA, USA, August 13-15, 2025},
  pages        = {489--507},
  publisher    = {{USENIX} Association},
  year         = {2025}
}

@inproceedings{8,
  author       = {Fengyu Liu and
                  Yuan Zhang and
                  Jiaqi Luo and
                  Jiarun Dai and
                  Tian Chen and
                  Letian Yuan and
                  Zhengmin Yu and
                  Youkun Shi and
                  Ke Li and
                  Chengyuan Zhou and
                  Hao Chen and
                  Min Yang},
  editor       = {Lujo Bauer and
                  Giancarlo Pellegrino},
  title        = {Make Agent Defeat Agent: Automatic Detection of Taint-Style Vulnerabilities
                  in LLM-based Agents},
  booktitle    = {34th {USENIX} Security Symposium, {USENIX} Security 2025, Seattle,
                  WA, USA, August 13-15, 2025},
  pages        = {3767--3786},
  publisher    = {{USENIX} Association},
  year         = {2025}
}

@inproceedings{12,
  author       = {Niklas Risse and
                  Marcel B{\"{o}}hme},
  editor       = {Davide Balzarotti and
                  Wenyuan Xu},
  title        = {Uncovering the Limits of Machine Learning for Automatic Vulnerability
                  Detection},
  booktitle    = {33rd {USENIX} Security Symposium, {USENIX} Security 2024, Philadelphia,
                  PA, USA, August 14-16, 2024},
  publisher    = {{USENIX} Association},
  year         = {2024}
}

@inproceedings{15,
  author       = {Zhongxin Liu and
                  Zhijie Tang and
                  Junwei Zhang and
                  Xin Xia and
                  Xiaohu Yang},
  title        = {Pre-training by Predicting Program Dependencies for Vulnerability
                  Analysis Tasks},
  booktitle    = {Proceedings of the 46th {IEEE/ACM} International Conference on Software
                  Engineering, {ICSE} 2024, Lisbon, Portugal, April 14-20, 2024},
  pages        = {151:1--151:13},
  publisher    = {{ACM}},
  year         = {2024}
}

@article{16,
  author       = {Jacob A. Harer and
                  Louis Y. Kim and
                  Rebecca L. Russell and
                  Onur Ozdemir and
                  Leonard R. Kosta and
                  Akshay Rangamani and
                  Lei H. Hamilton and
                  Gabriel I. Centeno and
                  Jonathan R. Key and
                  Paul M. Ellingwood and
                  Marc W. McConley and
                  Jeffrey M. Opper and
                  Peter Chin and
                  Tomo Lazovich},
  title        = {Automated software vulnerability detection with machine learning},
  journal      = {CoRR},
  volume       = {abs/1803.04497},
  year         = {2018}
}

@inproceedings{20,
  author       = {Zhen Li and
                  Deqing Zou and
                  Shouhuai Xu and
                  Xinyu Ou and
                  Hai Jin and
                  Sujuan Wang and
                  Zhijun Deng and
                  Yuyi Zhong},
  title        = {VulDeePecker: {A} Deep Learning-Based System for Vulnerability Detection},
  booktitle    = {25th Annual Network and Distributed System Security Symposium, {NDSS}
                  2018, San Diego, California, USA, February 18-21, 2018},
  publisher    = {The Internet Society},
  year         = {2018}
}

@inproceedings{21,
  author       = {Yizheng Chen and
                  Zhoujie Ding and
                  Lamya Alowain and
                  Xinyun Chen and
                  David A. Wagner},
  title        = {DiverseVul: {A} New Vulnerable Source Code Dataset for Deep Learning
                  Based Vulnerability Detection},
  booktitle    = {Proceedings of the 26th International Symposium on Research in Attacks,
                  Intrusions and Defenses, {RAID} 2023, Hong Kong, China, October 16-18,
                  2023},
  pages        = {654--668},
  publisher    = {{ACM}},
  year         = {2023}
}

@inproceedings{22,
  author       = {Rafael Rafailov and
                  Archit Sharma and
                  Eric Mitchell and
                  Christopher D. Manning and
                  Stefano Ermon and
                  Chelsea Finn},
  editor       = {Alice Oh and
                  Tristan Naumann and
                  Amir Globerson and
                  Kate Saenko and
                  Moritz Hardt and
                  Sergey Levine},
  title        = {Direct Preference Optimization: Your Language Model is Secretly a
                  Reward Model},
  booktitle    = {Advances in Neural Information Processing Systems 36: Annual Conference
                  on Neural Information Processing Systems 2023, NeurIPS 2023, New Orleans,
                  LA, USA, December 10 - 16, 2023},
  year         = {2023}
}

@inproceedings{23,
  author       = {Jiwoo Hong and
                  Noah Lee and
                  James Thorne},
  editor       = {Yaser Al{-}Onaizan and
                  Mohit Bansal and
                  Yun{-}Nung Chen},
  title        = {{ORPO:} Monolithic Preference Optimization without Reference Model},
  booktitle    = {Proceedings of the 2024 Conference on Empirical Methods in Natural
                  Language Processing, {EMNLP} 2024, Miami, FL, USA, November 12-16,
                  2024},
  pages        = {11170--11189},
  publisher    = {Association for Computational Linguistics},
  year         = {2024}
}

@article{24,
  author       = {Zhihong Shao and
                  Peiyi Wang and
                  Qihao Zhu and
                  Runxin Xu and
                  Junxiao Song and
                  Mingchuan Zhang and
                  Y. K. Li and
                  Y. Wu and
                  Daya Guo},
  title        = {DeepSeekMath: Pushing the Limits of Mathematical Reasoning in Open
                  Language Models},
  journal      = {CoRR},
  volume       = {abs/2402.03300},
  year         = {2024}
}

@article{25,
  author       = {DeepSeek{-}AI},
  title        = {DeepSeek-R1: Incentivizing Reasoning Capability in LLMs via Reinforcement
                  Learning},
  journal      = {CoRR},
  volume       = {abs/2501.12948},
  year         = {2025}
}

@inproceedings{26,
  author       = {Jingxuan He and
                  Martin T. Vechev},
  editor       = {Weizhi Meng and
                  Christian Damsgaard Jensen and
                  Cas Cremers and
                  Engin Kirda},
  title        = {Large Language Models for Code: Security Hardening and Adversarial
                  Testing},
  booktitle    = {Proceedings of the 2023 {ACM} {SIGSAC} Conference on Computer and
                  Communications Security, {CCS} 2023, Copenhagen, Denmark, November
                  26-30, 2023},
  pages        = {1865--1879},
  publisher    = {{ACM}},
  year         = {2023}
}

@article{28,
  author       = {Yue Li and
                  Xiao Li and
                  Hao Wu and
                  Minghui Xu and
                  Yue Zhang and
                  Xiuzhen Cheng and
                  Fengyuan Xu and
                  Sheng Zhong},
  title        = {Everything You Wanted to Know About LLM-based Vulnerability Detection
                  But Were Afraid to Ask},
  journal      = {CoRR},
  volume       = {abs/2504.13474},
  year         = {2025}
}

@article{29,
  author       = {Marco Simoni and
                  Aleksandar Fontana and
                  Giulio Rossolini and
                  Andrea Saracino},
  title        = {Improving {LLM} Reasoning for Vulnerability Detection via Group Relative
                  Policy Optimization},
  journal      = {CoRR},
  volume       = {abs/2507.03051},
  year         = {2025}
}

@article{30,
  author       = {Md Basim Uddin Ahmed and
                  Nima Shiri Harzevili and
                  Jiho Shin and
                  Hung Viet Pham and
                  Song Wang},
  title        = {SecVulEval: Benchmarking LLMs for Real-World {C/C++} Vulnerability
                  Detection},
  journal      = {CoRR},
  volume       = {abs/2505.19828},
  year         = {2025}
}

@inproceedings{31,
  author       = {Fabian Yamaguchi and
                  Nico Golde and
                  Daniel Arp and
                  Konrad Rieck},
  title        = {Modeling and Discovering Vulnerabilities with Code Property Graphs},
  booktitle    = {2014 {IEEE} Symposium on Security and Privacy, {SP} 2014, Berkeley,
                  CA, USA, May 18-21, 2014},
  pages        = {590--604},
  publisher    = {{IEEE} Computer Society},
  year         = {2014}
}

@inproceedings{32,
  author       = {Chenyuan Zhang and
                  Hao Liu and
                  Jiutian Zeng and
                  Kejing Yang and
                  Yuhong Li and
                  Hui Li},
  title        = {Prompt-Enhanced Software Vulnerability Detection Using ChatGPT},
  booktitle    = {Proceedings of the 2024 {IEEE/ACM} 46th International Conference on
                  Software Engineering: Companion Proceedings, {ICSE} Companion 2024,
                  Lisbon, Portugal, April 14-20, 2024},
  pages        = {276--277},
  publisher    = {{ACM}},
  year         = {2024}
}

@article{33,
  author       = {Youpeng Li and
                  Kartik Joshi and
                  Xinda Wang and
                  Eric Wong},
  title        = {{MAVUL:} Multi-Agent Vulnerability Detection via Contextual Reasoning
                  and Interactive Refinement},
  journal      = {CoRR},
  volume       = {abs/2510.00317},
  year         = {2025}
}

@article{34,
  author       = {Ratnadira Widyasari and
                  Martin Weyssow and
                  Ivana Clairine Irsan and
                  Han Wei Ang and
                  Frank Liauw and
                  Eng Lieh Ouh and
                  Lwin Khin Shar and
                  Hong Jin Kang and
                  David Lo},
  title        = {Let the Trial Begin: {A} Mock-Court Approach to Vulnerability Detection
                  using LLM-Based Agents},
  journal      = {CoRR},
  volume       = {abs/2505.10961},
  year         = {2025}
}

@inproceedings{35,
  author       = {Sihao Hu and
                  Tiansheng Huang and
                  Fatih Ilhan and
                  Selim Furkan Tekin and
                  Ling Liu},
  title        = {Large Language Model-Powered Smart Contract Vulnerability Detection:
                  New Perspectives},
  booktitle    = {5th {IEEE} International Conference on Trust, Privacy and Security
                  in Intelligent Systems and Applications, {TPS-ISA} 2023, Atlanta,
                  GA, USA, November 1-4, 2023},
  pages        = {297--306},
  publisher    = {{IEEE}},
  year         = {2023}
}

@inproceedings{36,
  author       = {Xin{-}Cheng Wen and
                  Yijun Yang and
                  Cuiyun Gao and
                  Yang Xiao and
                  Deheng Ye},
  editor       = {Wanxiang Che and
                  Joyce Nabende and
                  Ekaterina Shutova and
                  Mohammad Taher Pilehvar},
  title        = {Boosting Vulnerability Detection of LLMs via Curriculum Preference
                  Optimization with Synthetic Reasoning Data},
  booktitle    = {Findings of the Association for Computational Linguistics, {ACL} 2025,
                  Vienna, Austria, July 27 - August 1, 2025},
  pages        = {8935--8949},
  publisher    = {Association for Computational Linguistics},
  year         = {2025}
}

@article{37,
  author       = {Martin Weyssow and
                  Chengran Yang and
                  Junkai Chen and
                  Yikun Li and
                  Huihui Huang and
                  Ratnadira Widyasari and
                  Han Wei Ang and
                  Frank Liauw and
                  Eng Lieh Ouh and
                  Lwin Khin Shar and
                  David Lo},
  title        = {R2Vul: Learning to Reason about Software Vulnerabilities with Reinforcement
                  Learning and Structured Reasoning Distillation},
  journal      = {CoRR},
  volume       = {abs/2504.04699},
  year         = {2025}
}

@article{38,
  author       = {Jieh{-}Sheng Lee},
  title        = {InstructPatentGPT: training patent language models to follow instructions
                  with human feedback},
  journal      = {Artif. Intell. Law},
  volume       = {33},
  number       = {3},
  pages        = {739--782},
  year         = {2025}
}

@article{39,
  author       = {Aidan Z. H. Yang and
                  Haoye Tian and
                  He Ye and
                  Ruben Martins and
                  Claire {Le Goues}},
  title        = {Security Vulnerability Detection with Multitask Self-Instructed Fine-Tuning
                  of Large Language Models},
  journal      = {CoRR},
  volume       = {abs/2406.05892},
  year         = {2024}
}

@inproceedings{40,
  author       = {Yongshuai Liu and
                  Jiaxin Ding and
                  Xin Liu},
  title        = {{IPO:} Interior-Point Policy Optimization under Constraints},
  booktitle    = {The Thirty-Fourth {AAAI} Conference on Artificial Intelligence, {AAAI}
                  2020, The Thirty-Second Innovative Applications of Artificial Intelligence
                  Conference, {IAAI} 2020, The Tenth {AAAI} Symposium on Educational
                  Advances in Artificial Intelligence, {EAAI} 2020, New York, NY, USA,
                  February 7-12, 2020},
  pages        = {4940--4947},
  publisher    = {{AAAI} Press},
  year         = {2020}
}

@misc{nvd,
key = {National Vulnerability Database (NVD)},
howpublished = {\url{https://nvd.nist.gov/}},
year    = {2025},
}

@article{45,
  title={VulnLLM-R: Specialized Reasoning LLM with Agent Scaffold for Vulnerability Detection},
  author={Nie, Yuzhou and Li, Hongwei and Guo, Chengquan and Jiang, Ruizhe and Wang, Zhun and Li, Bo and Song, Dawn and Guo, Wenbo},
  journal={arXiv preprint arXiv:2512.07533},
  year={2025}
}

@inproceedings{46,
  author    = {Summers, Alec and Madden, Chris},
  title     = {Vulnerability Root Cause Mapping with {CWE}: Challenges, Solutions, and Insights from Grounded {LLM}-based Analysis},
  booktitle = {Proceedings of FIRST VulnCon 2025},
  year      = {2025},
  month     = {April},
  address   = {Scottsdale, AZ, USA},
  publisher = {Forum of Incident Response and Security Teams (FIRST)},
}

@inproceedings{50,
  author       = {Alperen Yildiz and
                  Sin G. Teo and
                  Yiling Lou and
                  Yebo Feng and
                  Chong Wang and
                  Dinil Mon Divakaran},
  title        = {Benchmarking LLMs and LLM-based Agents in Practical Vulnerability
                  Detection for Code Repositories},
  booktitle    = {Proceedings of the 63rd Annual Meeting of the Association for Computational
                  Linguistics (Volume 1: Long Papers), {ACL} 2025, Vienna, Austria,
                  July 27 - August 1, 2025},
  pages        = {30848--30865},
  publisher    = {Association for Computational Linguistics},
  year         = {2025},
}

@inproceedings{51,
  author       = {Guru Prasad Bhandari and
                  Amara Naseer and
                  Leon Moonen},
  title        = {CVEfixes: automated collection of vulnerabilities and their fixes
                  from open-source software},
  booktitle    = {{PROMISE} '21: 17th International Conference on Predictive Models
                  and Data Analytics in Software Engineering, Athens Greece, August
                  19-20, 2021},
  pages        = {30--39},
  publisher    = {{ACM}},
  year         = {2021}
}

@inproceedings{52,
  author       = {Jiahao Fan and
                  Yi Li and
                  Shaohua Wang and
                  Tien N. Nguyen},
  title        = {A {C/C++} Code Vulnerability Dataset with Code Changes and {CVE} Summaries},
  booktitle    = {{MSR} '20: 17th International Conference on Mining Software Repositories,
                  Seoul, Republic of Korea, 29-30 June, 2020},
  pages        = {508--512},
  publisher    = {{ACM}},
  year         = {2020}
}

@article{53,
  author       = {Yikun Li and
                  Ting Zhang and
                  Ratnadira Widyasari and
                  Yan Naing Tun and
                  Huu Hung Nguyen and
                  Tan Bui and
                  Ivana Clairine Irsan and
                  Yiran Cheng and
                  Xiang Lan and
                  Han Wei Ang and
                  Frank Liauw and
                  Martin Weyssow and
                  Hong Jin Kang and
                  Eng Lieh Ouh and
                  Lwin Khin Shar and
                  David Lo},
  title        = {CleanVul: Automatic Function-Level Vulnerability Detection in Code
                  Commits Using {LLM} Heuristics},
  journal      = {CoRR},
  volume       = {abs/2411.17274},
  year         = {2024}
}

@inproceedings{54,
  author       = {Chao Ni and
                  Liyu Shen and
                  Xiaohu Yang and
                  Yan Zhu and
                  Shaohua Wang},
  title        = {MegaVul: {A} {C/C++} Vulnerability Dataset with Comprehensive Code
                  Representations},
  booktitle    = {21st {IEEE/ACM} International Conference on Mining Software Repositories,
                  {MSR} 2024, Lisbon, Portugal, April 15-16, 2024},
  pages        = {738--742},
  publisher    = {{ACM}},
  year         = {2024}
}

@inproceedings{55,
  author       = {Xin Zhou and
                  Kisub Kim and
                  Bowen Xu and
                  DongGyun Han and
                  David Lo},
  title        = {Out of Sight, Out of Mind: Better Automatic Vulnerability Repair by
                  Broadening Input Ranges and Sources},
  booktitle    = {Proceedings of the 46th {IEEE/ACM} International Conference on Software
                  Engineering, {ICSE} 2024, Lisbon, Portugal, April 14-20, 2024},
  pages        = {88:1--88:13},
  publisher    = {{ACM}},
  year         = {2024}
}

@article{56,
  author       = {Yu Yang and
                  Yuzhou Nie and
                  Zhun Wang and
                  Yuheng Tang and
                  Wenbo Guo and
                  Bo Li and
                  Dawn Song},
  title        = {SecCodePLT: {A} Unified Platform for Evaluating the Security of Code
                  GenAI},
  journal      = {CoRR},
  volume       = {abs/2410.11096},
  year         = {2024}
}

@misc{57,
  title = {Clang: a C language family frontend for LLVM},
  howpublished = {\url{https://clang.llvm.org/}},
  year = {2026}
}

@misc{58,
  author       = {Anthropic},
  title        = {Claude Code},
  year         = {2026},
  howpublished = {\url{https://claude.ai/code}},
  note         = {AI coding assistant}
}

\appendix
\subsection{Comparative Analysis of Static Analysis Tools in VD}

\begin{table}[h]
\caption{Performance Comparison with Static Analysis Tools}
\label{tab:sast}
\centering
\begin{threeparttable}
\resizebox{0.8\columnwidth}{!}{
\renewcommand{\arraystretch}{1}
\begin{tabular}{ccccccc}
\toprule
\textbf{SAST Tool} & \multicolumn{3}{c}{\textbf{Binary Detection}} & \multicolumn{3}{c}{\textbf{Root Cause Detection}} \\ \midrule
                   & TP           & FN           & Recall          & TP            & FN             & Recall           \\ \midrule
CodeQL             & 6            & 978          & 0.61            & 1             & 983            & 0.10             \\
Cppcheck           & 63           & 921          & 6.40            & 1             & 983            & 0.10             \\
Flawfinder         & 204          & 780          & 20.73           & 16            & 968            & 1.63             \\
Joern Scan         & 89           & 895          & 9.04            & 5             & 979            & 0.51             \\ \bottomrule
\end{tabular}
}
\end{threeparttable}
\end{table}

To further understand the limitations of traditional static analysis tools in context-aware VD, we evaluate four representative SAST tools, including CodeQL, Cppcheck, Flawfinder, and Joern Scan, on the ContextVul test set. Following the hierarchical correctness evaluation introduced in Section~\ref{sec:sankey}, we report results under both binary detection and root-cause-level detection settings.

As shown in Table~\ref{tab:sast}, all evaluated SAST tools exhibit substantial performance degradation under root-cause-level evaluation compared to binary detection. This observation further validates our conclusion that coarse-grained binary evaluation can significantly overestimate practical vulnerability understanding capability.

Among the evaluated tools, CodeQL adopts the most conservative detection strategy, achieving extremely low recall under both evaluation settings. Although CodeQL can generate detailed data-flow traces and semantically meaningful warnings, its effectiveness heavily depends on complete project compilation environments and accurate build configurations. Consequently, it performs poorly when analyzing isolated vulnerability snippets or incomplete repository contexts, which are common in real-world VD benchmarks.

Cppcheck and Joern Scan also exhibit severe drops in true positives under root-cause-level evaluation. For example, Cppcheck detects 63 vulnerable samples under binary detection but correctly identifies the actual vulnerability root cause for only one sample. Our analysis shows that many of its warnings are triggered by generic syntax heuristics or shallow buffer checks rather than the target vulnerabilities themselves. Similarly, Joern Scan frequently produces detections based on coarse structural patterns such as loop-copy behaviors, resulting in poor vulnerability-specific reasoning capability.

Flawfinder achieves the strongest performance among all evaluated SAST tools, primarily because it relies on lexical matching over dangerous API functions such as \texttt{memcpy} and \texttt{strcpy}. However, despite being effective at identifying potentially risky sinks, Flawfinder still struggles to distinguish vulnerable code from patched code when both contain similar API usage patterns.

\noindent\textbf{Limitations Compared to VULPO.}
Compared with traditional SAST tools, VULPO demonstrates several important advantages for context-aware VD. First, SAST tools generally rely on complete compilation environments or handcrafted heuristic rules, limiting their effectiveness in incomplete or non-compilable repository settings. In contrast, VULPO adaptively reasons over repository-level contextual information without requiring successful project compilation.

Second, existing SAST tools exhibit limited capability in distinguishing vulnerable code from subtle patched variants, particularly when both share highly similar syntactic structures. Through the cold-start reasoning initialization and difficulty-aware RL optimization, VULPO learns to differentiate subtle vulnerability semantics and patch behaviors more effectively.

Finally, while traditional SAST tools typically output only technical warnings or line-level alerts, VULPO generates structured and context-aware vulnerability reasoning that explains the underlying root cause and relevant program semantics. Such explainable vulnerability analysis not only improves VD performance but also provides valuable prior knowledge for downstream tasks such as vulnerability patching, security auditing, and remediation assistance.

\subsection{LLM-Judge Reward Evaluation Protocol}\label{sec:specification}

\subsubsection{Vulnerable Version}

Figure~\ref{box:sample_judge_vul} presents the process-aware evaluation prompt used for vulnerable samples. The judge LLM evaluates whether the detector correctly identifies the target vulnerability by comparing the generated analysis against a predefined vulnerability checklist derived from CVE descriptions, patch commit messages, and vulnerability-relevant code semantics. To provide fine-grained supervision, the evaluation is decomposed into three complementary dimensions: outcome correctness, vulnerability-relevant localization, and causal reasoning quality.

Specifically, the \texttt{Verdict} dimension evaluates whether the detector correctly identifies the target vulnerability type regardless of any additional unrelated findings. The \texttt{Evidence} dimension measures whether the detector accurately localizes the vulnerability-relevant code regions specified in the checklist. Finally, the \texttt{Reasoning} dimension evaluates whether the detector correctly explains both the root cause and technical consequence of the vulnerability. The judge outputs a structured JSON object containing both categorical judgments and concise justifications for each dimension.

\begin{figure}[!t]
\begin{tcolorbox}[boxrule=1pt, left=1mm, right=1mm, top=1mm, bottom=1mm, fontupper=\scriptsize]
\texttt{<|system|>} You are to act as a meticulous and impartial Code Security Expert and Evaluator.\\
\texttt{<|user|>}\\
\textbf{1. Goal}\\
Your task is to evaluate a vulnerability analysis for a code sample known to be \texttt{HAS\_VUL} (vulnerable). Verify if the detector successfully identified the specific target vulnerability described in the Checklist.\\

\textbf{2. Scoring Rubric}\\
Dimension 1: Verdict (Strict Binary)\\
- \texttt{CORRECT:} The analysis identifies the specific vulnerability type requested in the checklist (e.g., "Buffer Overflow"). Ignore any extra/unrelated vulnerabilities listed by the detector. As long as the target is found, it is true.\\
- \texttt{INCORRECT:} The analysis fails to mention the target vulnerability type.\\

Dimension 2: Evidence (Insecure Code Snippet)\\
- \texttt{CORRECT:} The analysis quotes the exact code snippet (or a semantically identical variation) requested in the checklist.\\
- \texttt{PARTIALLY CORRECT:} The analysis cites the correct variable name or surrounding logic, but does not quote the specific snippet exactly. Or, it quotes a huge block of code that contains the snippet but lacks precision.\\
- \texttt{INCORRECT:} No specific evidence or incorrect code cited.\\

Dimension 3: Reasoning (Causal Vulnerability Analysis)\\
- \texttt{CORRECT:} The analysis correctly explains both the root cause (e.g., "missing check") and the consequence (e.g., "overflow").\\
- \texttt{PARTIALLY CORRECT:} Explains the general issue (e.g., "unsafe copy") but misses technical details.\\
- \texttt{INCORRECT:} Incorrect or generic reasoning.\\

\textbf{3. Input}\\
The Ground Truth Checklist (Target: Vulnerable):\\
\{CHECKLIST\}\\

The Candidate Analysis:\\
\{ANALYSIS\}\\

\textbf{4. Output Format (Json Only)}\\
\verb|```|json\\
\{\\
\hspace*{1em}"Verdict\_Recall": \{\\
\hspace*{2em}"justification": "$<$Your brief reason$>$",\\
\hspace*{2em}"option": $<$choose from ["\texttt{CORRECT}", "\texttt{INCORRECT}"]$>$\\
\hspace*{1em}\},\\
\hspace*{1em}"Evidence\_Insecure\_Code": \{\\
\hspace*{2em}"justification": "$<$Your brief reason$>$",\\
\hspace*{2em}"option": $<$choose from ["\texttt{CORRECT}", "\texttt{PARTIALLY CORRECT}", "\texttt{INCORRECT}"]$>$\\
\hspace*{1em}\},\\
\hspace*{1em}"Reasoning\_Mechanism": \{\\
\hspace*{2em}"justification": "$<$Your brief reason$>$",\\
\hspace*{2em}"option": $<$choose from ["\texttt{CORRECT}", "\texttt{PARTIALLY CORRECT}", "\texttt{INCORRECT}"]$>$\\
\hspace*{1em}\}\\
\}\\
\verb|```|
\end{tcolorbox}
\caption{Process-Aware LLM-Judge Prompt for Vulnerable Samples}
\label{box:sample_judge_vul}
\end{figure}

\subsubsection{Patched Version}

Figure~\ref{box:sample_judge_fix} presents the evaluation prompt for patched samples, which focuses on assessing whether the detector correctly recognizes the absence of the target vulnerability after patching. Unlike vulnerable-sample evaluation, this protocol emphasizes the model’s ability to identify safeguard logic introduced by patches and avoid hallucinating previously fixed vulnerabilities.

The \texttt{Verdict} dimension evaluates whether the detector correctly concludes that the specific target vulnerability has been resolved. The \texttt{Evidence} dimension measures whether the detector accurately identifies safeguard code introduced by the patch, such as sanitizers, boundary checks, or validation logic. Finally, the \texttt{Reasoning} dimension evaluates whether the detector correctly explains why the patch successfully mitigates the original vulnerability. The resulting structured JSON output provides fine-grained supervision over both safety reasoning and false-positive avoidance.

\begin{figure}[!t]
\begin{tcolorbox}[boxrule=1pt, left=1mm, right=1mm, top=1mm, bottom=1mm, fontupper=\scriptsize]
\texttt{<|system|>} You are to act as a meticulous and impartial Code Security Expert and Evaluator.\\
\texttt{<|user|>}\\
\textbf{1. Goal}\\
Your task is to evaluate a vulnerability analysis for a code sample known to be \texttt{NO\_VUL} (vulnerability fixed). Verify if the detector correctly concludes the code is safe regarding the specific vulnerability.\\

\textbf{2. Scoring Rubric}\\
Dimension 1: Verdict (Strict Binary)\\
- \texttt{CORRECT:} The analysis says "\texttt{NO\_VUL}" or The analysis reports other vulnerabilities but does not list the specific fixed vulnerability mentioned in the checklist.\\
- \texttt{INCORRECT:} The analysis explicitly claims the specific target vulnerability (e.g., the one described in the checklist) still exists.\\

Dimension 2: Evidence (Safeguard Snippet)\\
- \texttt{CORRECT:} The analysis quotes the exact safeguard code (e.g., the new check/sanitizer) requested in the checklist.\\
- \texttt{PARTIALLY CORRECT:} The analysis cites the correct variable name or surrounding logic, but does not quote the specific snippet exactly. Or, it quotes a huge block of code that contains the snippet but lacks precision.\\
- \texttt{INCORRECT:} No specific evidence or incorrect code cited.\\

Dimension 3: Reasoning (Causal Safety Analysis)\\
- \texttt{CORRECT:} Explains why the code is safe (e.g., "The new check prevents the overflow").\\
- \texttt{PARTIALLY CORRECT:} Vague acknowledgment of safety without specific logic.\\
- \texttt{INCORRECT:} Incorrect logic or claims the code is unsafe.\\

\textbf{3. Input}\\
The Ground Truth Checklist (Target: Safe/Fixed):\\
\{CHECKLIST\}\\

The Candidate Analysis:\\
\{ANALYSIS\}\\

\textbf{4. Output Format (Json Only)}\\
\verb|```|json\\
\{\\
\hspace*{1em}"Verdict\_Absence\_of\_Specific\_Vuln": \{\\
\hspace*{2em}"justification": "$<$Your brief reason$>$",\\
\hspace*{2em}"option": $<$choose from ["\texttt{CORRECT}", "\texttt{INCORRECT}"]$>$\\
\hspace*{1em}\},\\
\hspace*{1em}"Evidence\_Safeguard\_Code": \{\\
\hspace*{2em}"justification": "$<$Your brief reason$>$",\\
\hspace*{2em}"option": $<$choose from ["\texttt{CORRECT}", "\texttt{PARTIALLY CORRECT}", "\texttt{INCORRECT}"]$>$\\
\hspace*{1em}\},\\
\hspace*{1em}"Reasoning\_Resolution": \{\\
\hspace*{2em}"justification": "$<$Your brief reason$>$",\\
\hspace*{2em}"option": $<$choose from ["\texttt{CORRECT}", "\texttt{PARTIALLY CORRECT}", "\texttt{INCORRECT}"]$>$\\
\hspace*{1em}\}\\
\}\\
\verb|```|
\end{tcolorbox}
\caption{Process-Aware LLM-Judge Prompt for Patched Samples}
\label{box:sample_judge_fix}
\end{figure}

This dual-protocol design enables the judge LLM to separately evaluate vulnerability reasoning and safety reasoning, thereby providing faithful process-aware supervision for both vulnerable and patched code analysis. The categorical judgments are subsequently mapped to numerical rewards according to the reward aggregation rules described in Section~\ref{sec:reward_judge}.

\subsection{Case Study: Context-Aware Reasoning for Use-After-Free Detection}

\subsubsection{Vulnerable Version}

\begin{figure}[!t]
\begin{tcolorbox}[boxrule=1pt, left=1mm, right=1mm, top=1mm, bottom=1mm, fontupper=\scriptsize]
\texttt{<|system|>} You are a vulnerability detection expert specializing in identifying security flaws in C/C++ code, with a focus on Common Weakness Enumeration (CWE) standards. You provide precise, evidence-based analysis without speculation, and clearly label any vulnerabilities you detect.\\
\texttt{<|user|>}\\
Your task is to evaluate whether the following C/C++ code contains any security vulnerabilities.
\\
You will be provided with two sections:\\
1. Context: Relevant code such as includes, type definitions, global variables, macros, and definitions of any functions called within the target function.\\
2. Code: The target function to analyze.
\\\\
Use all available information to analyze the function step by step.\\
If the target function alone is insufficient to determine whether a vulnerability exists, refer to the Context section before making a judgment.\\
Do not assume vulnerabilities — only report what is supported by the code and context.\\
\\
In your final response, list all detected vulnerabilities and CWE identifiers if applicable.\\
Conclude with one of the following indicators on a new line:\\
- HAS\_VUL — if any vulnerabilities are found \\
- NO\_VUL — if no vulnerabilities are found \\
\\
\verb|```|Context\\
Includes:\\
\#include $<$linux/dmi.h$>$\\
\#include $<$linux/hid.h$>$\\
\#include $<$linux/module.h$>$\\
\#include $<$linux/platform\_data/x86/asus-wmi.h$>$\\
\#include $<$linux/input/mt.h$>$\\
\#include $<$linux/usb.h$>$\\
\#include $<$linux/power\_supply.h$>$\\
\#include "hid-ids.h"\\
\\
Type Definitions:\\
- led\_classdev: struct led\_classdev\\
\\
Called Methods:\\
- spin\_lock\_irqsave: static inline void spin\_lock\_irqsave(spinlock\_t *lock, unsigned long f \{spin\_lock(lock);\}\\
- spin\_unlock\_irqrestore: static inline void spin\_unlock\_irqrestore(spinlock\_t *lock, unsigned long f) \{spin\_unlock(lock);\}\\
- schedule\_work: static inline bool schedule\_work(struct work\_struct *work)
\{
        return queue\_work(system\_wq, work);
\}\\
- spin\_lock: static inline void spin\_lock(spinlock\_t *lock) \{int ret = pthread\_spin\_lock(lock); assert(!ret); \}\\
- spin\_unlock: static inline void spin\_unlock(spinlock\_t *lock) \{int ret = pthread\_spin\_unlock(lock); assert(!ret);\}\\
- queue\_work: static inline bool queue\_work(struct workqueue\_struct *wq, struct work\_struct *work) \{return queue\_work\_on(WORK\_CPU\_UNBOUND, wq, work);\}  \\
\verb|```|\\
\\
\verb|```|Code\\
File: drivers/hid/hid-asus.c\\
Method: asus\_kbd\_backlight\_set\\
----------------------------------------\\
static void asus\_kbd\_backlight\_set(struct led\_classdev *led\_cdev, enum led\_brightness brightness)\\
\{\\
        struct asus\_kbd\_leds *led = container\_of(led\_cdev, struct asus\_kbd\_leds,
                                                 cdev);\\
        unsigned long flags;\\
        spin\_lock\_irqsave(\&led-$>$lock, flags);\\
        led-$>$brightness = brightness;\\
        spin\_unlock\_irqrestore(\&led-$>$lock, flags);\\
        \textbf{\textcolor{red}{schedule\_work(\&led-$>$work);}}\\
\}\\
\verb|```|\\
\\
Analyze the code now.
\end{tcolorbox}
\caption{CVE-2023-1079 (Use-After-Free): Vulnerable Code Sample as LLM Detection Input}
\label{box:case_vul_input}
\end{figure}

\begin{figure}[!t]
\begin{tcolorbox}[boxrule=1pt, left=1mm, right=1mm, top=1mm, bottom=1mm, fontupper=\scriptsize]
$<$think$>$\\
We are given the context and the code for the function `asus\_kbd\_backlight\_set`. The function is part of a HID driver for ASUS keyboards.\\
\\
 Steps:\\
 1. The function takes a `struct led\_classdev *led\_cdev` and an `enum led\_brightness brightness`.\\
 2. It uses `container\_of` to get a pointer to the containing `struct asus\_kbd\_leds` from the `led\_cdev`.\\
 3. It then locks a spinlock (using `spin\_lock\_irqsave`) to protect the `brightness` field of the `asus\_kbd\_leds` structure.\\
 4. It sets `led-$>$brightness` to the provided `brightness`.\\
 5. It unlocks the spinlock (using `spin\_unlock\_irqrestore`).\\
 6. Finally, it schedules work (using `schedule\_work`) for the `led-$>$work` work item.\\
\\
 Now, let's look at the context:\\
 The context includes several spinlock functions and `schedule\_work`.\\
 ...\\
 \textbf{\textcolor{red}{However, the main concern is the use-after-free. How can the structure be freed?  The function `asus\_kbd\_backlight\_set` is called when the brightness is changed. The LED subsystem will call this function. If the device is being removed, the LED subsystem might unregister the LED and then free the structure. But note that the work is scheduled and might run after the structure is freed.
 In the Linux kernel, when a device is being removed, the driver should ensure that all pending work is canceled. However, we don't see any cancellation of the work in this function. Therefore, if the device is removed, the work might still be pending and run after the structure is freed. This is a vulnerability. CWE-416: Use-After-Free}}\\
...\\
 How to fix? The driver should cancel the work when the device is being removed. But in the given code, there is no such cancellation. Therefore, we have a use-after-free.\\
$<$/think$>$\\
\\
\#\#\# Analysis of `asus\_kbd\_backlight\_set`\\
\\
\#\#\#\# Step-by-Step Evaluation:\\
1. **Locking Mechanism**:  \\
   The function uses `spin\_lock\_irqsave` and `spin\_unlock\_irqrestore` to protect the `led-$>$brightness` assignment. This ensures atomicity during the write operation, preventing race conditions (e.g., concurrent access to `led-$>$brightness`). The locking is correctly paired and releases the spinlock before proceeding.\\
\textbf{\textcolor{red}{2. **Work Scheduling**:}}  \\
   \textbf{\textcolor{red}{After updating `led-$>$brightness`, the function calls `schedule\_work(\&led-$>$work)`. This schedules an asynchronous work item to execute later. However, **no validation is performed** to ensure `led` (or its members, including `led-$>$work`) remain valid after the spinlock is released. If the device is being unregistered or the `asus\_kbd\_leds` structure is freed before the work executes, the work function may access invalid memory.}}\\
3. **Use-After-Free Risk**:  \\
   The core vulnerability arises from scheduling work (`schedule\_work`) without checking the lifetime of `led`. If the device is removed (e.g., via `platform\_driver\_remove`), the `asus\_kbd\_leds` structure could be freed, making `led-$>$work` and `led-$>$brightness` dangling pointers. When the work executes, it will dereference these invalid pointers, leading to undefined behavior (e.g., crashes, data corruption).\\
\textbf{\textcolor{red}{4. **Contextual Evidence**:}}\\
   \textbf{\textcolor{red}{- The provided context shows `schedule\_work` is a kernel API that queues work asynchronously.}}\\
   \textbf{\textcolor{red}{- No mechanisms (e.g., work cancellation) are present to handle device removal during this operation.}}\\
   \textbf{\textcolor{red}{- The function lacks safeguards (e.g., reference counting) to ensure `led` remains valid after the spinlock is released.}}\\
\\
\#\#\#\# Identified Vulnerability:\\
- **CWE-416: Use-After-Free**  \\
  Scheduling work (`schedule\_work`) after modifying shared state (`led-$>$brightness`) without ensuring the object's lifetime is valid risks accessing freed memory. This violates kernel memory safety.\\
\\
\#\#\# Final Output:\\
Detected vulnerabilities:  \\
- Use-After-Free (CWE-416)\\
\\
HAS\_VUL
\end{tcolorbox}
\caption{VULPO Contextual Analysis of Vulnerable Code Sample: CVE-2023-1079 (Use-After-Free)}
\label{box:case_vul_output}
\end{figure}

Figure~\ref{box:case_vul_input} presents the vulnerable version of the target function \texttt{asus\_kbd\_backlight\_set}. The function updates LED brightness under a spinlock and subsequently invokes \texttt{schedule\_work(\&led->work)} after releasing the lock. Although the assignment to \texttt{led->brightness} is protected by a spinlock, the synchronization only guarantees atomic access to the field itself and does not ensure the lifetime safety of the \texttt{led} object across asynchronous execution contexts.

The core issue arises because \texttt{schedule\_work} introduces deferred asynchronous execution without any mechanism guaranteeing that the \texttt{struct asus\_kbd\_leds *led} object remains valid until the scheduled work completes. If the device is unplugged or removed concurrently, the LED structure may be freed while the queued work item is still pending. Consequently, the scheduled work function may dereference a dangling pointer, resulting in a classic use-after-free vulnerability (CWE-416). This issue is particularly dangerous in adversarial scenarios involving malicious USB devices that repeatedly trigger rapid connect and disconnect events.

\noindent\textbf{VULPO Contextual Analysis of the Vulnerability.}
As shown in Figure~\ref{box:case_vul_output}, VULPO correctly identifies the vulnerability by reasoning beyond local synchronization patterns. Rather than treating the existence of a spinlock as sufficient evidence of safety, VULPO analyzes the temporal separation between lock release and asynchronous execution. It recognizes that \texttt{schedule\_work} introduces a delayed execution context that is no longer protected by the original lock.

More importantly, VULPO reasons about object lifetime across execution contexts and correctly infers that the \texttt{led} structure may be freed before the scheduled work executes. The generated reasoning explicitly connects asynchronous work scheduling, device removal, and dangling pointer dereferencing, demonstrating a context-aware understanding of the root cause rather than simple pattern matching.

\noindent\textbf{Failure of Other SOTA LLMs.}
In contrast, several frontier reasoning LLMs incorrectly classify the vulnerable version as safe or misidentify the vulnerability type. Their analyses primarily rely on local synchronization patterns and incorrectly treat the presence of a spinlock as sufficient evidence of safety. As a result, they fail to reason about the temporal gap introduced by asynchronous work scheduling and overlook the object lifetime risks across execution contexts.

\subsubsection{Patched Version}

\begin{figure}[!t]
\begin{tcolorbox}[boxrule=1pt, left=1mm, right=1mm, top=1mm, bottom=1mm, fontupper=\scriptsize]
\texttt{<|system|>} You are a vulnerability detection expert specializing in identifying security flaws in C/C++ code, with a focus on Common Weakness Enumeration (CWE) standards. You provide precise, evidence-based analysis without speculation, and clearly label any vulnerabilities you detect.\\
\texttt{<|user|>}\\
Your task is to evaluate whether the following C/C++ code contains any security vulnerabilities.\\
\\
You will be provided with two sections:\\
1. Context: Relevant code such as includes, type definitions, global variables, macros, and definitions of any functions called within the target function.\\
2. Code: The target function to analyze.\\
\\
Use all available information to analyze the function step by step.\\
If the target function alone is insufficient to determine whether a vulnerability exists, refer to the Context section before making a judgment.\\
Do not assume vulnerabilities — only report what is supported by the code and context.\\
\\
In your final response, list all detected vulnerabilities and CWE identifiers if applicable.\\
Conclude with one of the following indicators on a new line:\\
- HAS\_VUL — if any vulnerabilities are found\\
- NO\_VUL — if no vulnerabilities are found\\
    \\
\verb|```|Context\\
Includes:\\
- \#include $<$linux/dmi.h$>$\\
- \#include $<$linux/hid.h$>$\\
- \#include $<$linux/module.h$>$\\
- \#include $<$linux/platform\_data/x86/asus-wmi.h$>$\\
- \#include $<$linux/input/mt.h$>$\\
- \#include $<$linux/usb.h$>$\\
- \#include $<$linux/power\_supply.h$>$\\
- \#include "hid-ids.h"\\
\\
Type Definitions:\\
- led\_classdev: struct led\_classdev\\
\\
Called Methods:\\
- spin\_lock\_irqsave: static inline void spin\_lock\_irqsave(spinlock\_t *lock, unsigned long f) \{spin\_lock(lock);\}\\
- spin\_unlock\_irqrestore: static inline void spin\_unlock\_irqrestore(spinlock\_t *lock, unsigned long f) \{spin\_unlock(lock);\}\\
- asus\_schedule\_work: static void asus\_schedule\_work(struct asus\_kbd\_leds *led)
\{
        unsigned long flags;
        spin\_lock\_irqsave(\&led-$>$lock, flags);
        if (!led-$>$removed)
                schedule\_work(\&led-$>$work);
        spin\_unlock\_irqrestore(\&led-$>$lock, flags);
\}\\
- spin\_lock: static inline void spin\_lock(spinlock\_t *lock) \{int ret = pthread\_spin\_lock(lock); assert(!ret); \}\\
- spin\_unlock: static inline void spin\_unlock(spinlock\_t *lock) \{int ret = pthread\_spin\_unlock(lock); assert(!ret);\}\\
- schedule\_work: static inline bool schedule\_work(struct work\_struct *work) \{return queue\_work(system\_wq, work);\}\\
- queue\_work: static inline bool queue\_work(struct workqueue\_struct *wq, struct work\_struct *work) \{return queue\_work\_on(WORK\_CPU\_UNBOUND, wq, work);\}\\
\verb|```|\\
\\
\verb|```|Code\\
File: drivers/hid/hid-asus.c\\
Method: asus\_kbd\_backlight\_set\\
----------------------------------------\\
static void asus\_kbd\_backlight\_set(struct led\_classdev *led\_cdev,
                                   enum led\_brightness brightness)\\
\{\\
        struct asus\_kbd\_leds *led = container\_of(led\_cdev, struct asus\_kbd\_leds,
                                                 cdev);\\
        unsigned long flags;\\
        spin\_lock\_irqsave(\&led-$>$lock, flags);\\
        led-$>$brightness = brightness;\\
        spin\_unlock\_irqrestore(\&led-$>$lock, flags);\\
        \textbf{\textcolor{Green}{asus\_schedule\_work(led)}};\\
\}\\
\verb|```|\\
\\
Analyze the code now.
\end{tcolorbox}
\caption{CVE-2023-1079 (Use-After-Free): Fixed Code Sample as LLM Detection Input}
\label{box:case_fix_input}
\end{figure}

\begin{figure}[!t]
\begin{tcolorbox}[boxrule=1pt, left=1mm, right=1mm, top=1mm, bottom=1mm, fontupper=\scriptsize]
$<$think$>$\\
We are given the context and the target function `asus\_kbd\_backlight\_set`.\\
\\
 Steps:\\
 1. The function takes a `led\_classdev` pointer and a brightness value.\\
 2. It uses `container\_of` to get a pointer to the enclosing `asus\_kbd\_leds` structure.\\
 3. It then acquires a spinlock with `spin\_lock\_irqsave` to protect the access to `led-$>$brightness`.\\
 4. It sets `led-$>$brightness` to the provided `brightness` value.\\
 5. It releases the spinlock with `spin\_unlock\_irqrestore`.\\
 6. It calls `asus\_schedule\_work(led)`.\\
\\
 Now, let's look at the provided context for called methods: We have definitions for spinlock functions and `schedule\_work`.\\
 ...\\
 The function `asus\_schedule\_work` is provided:\\
 ...\\
 Now, let's analyze for potential vulnerabilities:\\
 1. Double Fetch or TOCTOU (Time of Check, Time of Use) issues? ...\\
 2. Locking issues: The function uses a spinlock to protect the `brightness` assignment. This is correct. The function then calls `asus\_schedule\_work(led)`. The `asus\_schedule\_work` function also uses the same lock to check `led-$>$removed`. This is also correct.\\
 \textbf{\textcolor{Green}{3. What if `asus\_schedule\_work` is called while the device is being removed? The `asus\_schedule\_work` function checks `led-$>$removed` and only schedules work if it is false. So if the device is being removed, the work is not scheduled.}}\\
 4. However, note that the `asus\_kbd\_backlight\_set` function does not check if the device is still present. ... So it's safe.\\
 5. But note: the `asus\_kbd\_backlight\_set` function does not have any check for the state of the device. ... So it's safe.\\
 6. However, there is a potential issue: the `asus\_schedule\_work` function is called after releasing the lock. What if the device is removed after the lock is released but before `asus\_schedule\_work` is called? ... Therefore, it is safe.\\
...\\
 After careful analysis, I do not see any vulnerability in the provided code. However, note that the function `asus\_schedule\_work` is called after releasing the lock. If the device is removed after the lock is released, the `asus\_schedule\_work` function will still run and check the `removed` flag. But as explained, it will not schedule the work. So it's safe.\\
\\
 Therefore, no vulnerabilities are detected.\\
$<$/think$>$\\
\\
\#\#\# Analysis of `asus\_kbd\_backlight\_set`\\
\\
\#\#\#\# Step-by-Step Evaluation:\\
1. **Locking Mechanism**:\\
   - The function uses `spin\_lock\_irqsave` and `spin\_unlock\_irqrestore` to protect access to `led-$>$brightness`. This ensures atomicity when updating `led-$>$brightness`, preventing race conditions during the assignment.\\
\textbf{\textcolor{Green}{2. **Post-Lock Operation**:}}\\
   \textbf{\textcolor{Green}{- After releasing the spinlock, `asus\_schedule\_work(led)` is called. The provided `asus\_schedule\_work` function re-acquires the same spinlock and checks `led-$>$removed` before scheduling work. This ensures:}}\\
   \textbf{\textcolor{Green}{- The work is only scheduled if `led-$>$removed` is `false`.}}\\
   \textbf{\textcolor{Green}{- No operations proceed if the device is being removed, preventing use-after-free or invalid state access.}}\\
3. **No Vulnerabilities Detected**:\\
   - **No TOCTOU (Time-of-Check-Time-of-Use) Race**: The `asus\_schedule\_work` function re-checks `led-$>$removed` under the lock, ensuring safety even if the device state changes after the initial lock release.\\
   - **No Double Fetch**: The `brightness` value is set atomically under the lock, and subsequent work functions (not shown) would re-read it under the same lock.\\
   \textbf{\textcolor{Green}{- **No Invalid State Access**: The `asus\_schedule\_work` function explicitly guards against scheduling work when `led-$>$removed` is true, avoiding operations on invalid memory.}}\\
   - **No Out-of-Bounds Risks**: The `brightness` parameter is an `enum led\_brightness`, which is validated by the LED subsystem (caller responsibility), and the function does not perform additional checks.\\
\\
\#\#\#\# Conclusion:\\
The code correctly uses locking to protect shared data and safely handles device removal states. No security vulnerabilities are identified.\\
\\
**NO\_VUL**\\
\end{tcolorbox}
\caption{VULPO Contextual Analysis of Vulnerability-Fixed Code Sample: CVE-2023-1079 (Use-After-Free)}
\label{box:case_fix_output}
\end{figure}

Figure~\ref{box:case_fix_input} presents the patched version of the same function. Unlike the vulnerable implementation, the patched version introduces an explicit lifetime-aware scheduling invariant through the wrapper function \texttt{asus\_schedule\_work(led)}.

Specifically, instead of directly invoking \texttt{schedule\_work(\&led->work)}, the patched code re-acquires the same spinlock and checks the \texttt{led->removed} flag before scheduling any asynchronous work. This additional safeguard ensures that no new work item is queued once device removal has begun, thereby preventing access to freed memory.

\noindent\textbf{VULPO Contextual Analysis of the Vulnerability Fix.}
As shown in Figure~\ref{box:case_fix_output}, VULPO correctly recognizes that the patch fundamentally changes the security semantics of the code. Rather than relying solely on syntactic differences, VULPO performs inter-procedural reasoning over the implementation of \texttt{asus\_schedule\_work}.

The model correctly infers that the combination of (1) re-acquiring the spinlock and (2) validating the device removal state establishes a lifetime-aware scheduling discipline that eliminates the original use-after-free condition. Importantly, VULPO identifies the implicit safety invariant introduced by the patch: asynchronous work must not be scheduled once the device enters the removal state.

This example highlights that effective VD requires reasoning not only about explicit unsafe operations, but also about implicit temporal and lifetime guarantees introduced by patches.

Overall, this case study demonstrates that VULPO does not merely memorize vulnerability patterns or dangerous APIs. Instead, it learns to reason over asynchronous execution semantics, object lifetime constraints, and implicit safety invariants using repository-level contextual information.

\end{document}